\documentclass[useAMS,usenatbib]{mn2e}  

\def\lsim{~\rlap{$<$}{\lower 1.0ex\hbox{$\sim$}}}
\def\gsim{~\rlap{$>$}{\lower 1.0ex\hbox{$\sim$}}}

\def\e{$\pm$} 
\newcommand{\ltsima} {$\; \buildrel < \over \sim \;$} 
\newcommand{\simlt} {\lower.5ex\hbox{\ltsima}} 
\newcommand{\gtsima} {$\; \buildrel > \over \sim \;$} 
\newcommand{\simgt} {\lower.5ex\hbox{\gtsima}} 
\def\kms{km$\,$s$^{-\!1}$} 
\def\vsi{$v\: \sin i$} 
\def\rsun{R$_{\odot}$}

\def\msun{M$_{\odot}$}

\title[Rotational velocities in symbiotic stars - II]
{Rotational velocities of the giants in symbiotic stars: \\ II. Are S-type symbiotics 
synchronized? 
\thanks{based on observations obtained in ESO programs 073.D-0724A
              and 074.D-0114}} 
\author[Zamanov, Bode, Melo, et al. ] {
R. K. Zamanov$^{1,2}$  
\thanks{e-mail: rkz@astro.bas.bg; mfb@astro.livjm.ac.uk; andreja.gomboc@fmf.uni-lj.si}, 
M. F. Bode$^{2}$, 
C. H. F. Melo$^{3,4}$, R. Bachev$^{1}$, \\
\\ 
\LARGE 
{\rm A. Gomboc$^{2,5}$, I. K. Stateva$^{1}$, J. M. Porter$^{2}$, J. Pritchard$^{3}$ }\\
\\
$^1$ Institute of Astronomy, Bulgarian Academy of Sciences, 
       72 Tsarighradsko Shousse Blvd., 1784 Sofia, Bulgaria \\ 
$^2$ Astrophysics Research Institute, Liverpool John Moores University, 
       Twelve Quays House, Birkenhead, CH41 1LD, UK\\ 
$^3$ European Southern Observatory, Casilla 19001, Santiago 19, Chile \\
$^4$ Departamento de Astronom\'{\i}a, Universidad de Chile, Casilla 36-D, Santiago, Chile  \\
$^5$ Department of Physics, University of Ljubljana, 
       Jadranska 19, 6100 Ljubljana, Slovenia
}  
\begin{document} 
\date{Accepted . Received 2007 May 17; in original form 2006 March 17} 
 \pagerange{\pageref{firstpage}--\pageref{lastpage}} \pubyear{2005} 
 
\maketitle 
 
\label{firstpage} 
 
\begin{abstract} 
We have measured the projected rotational velocities (\vsi) of the mass donors for 
29 S-type symbiotic stars using high resolution spectroscopic observations 
and  the cross-correlation function (CCF) method. The results of the CCF
have been controlled with synthetic spectra.  
The typical rotational velocity of the K and M giants in S-type symbiotics
appeared to be $4.5<$\vsi$<11.7$ \kms.  
In a sub-sample of 16 S--type symbiotic stars 
(with known orbital periods and well measured \vsi)
15 have deviations from synchronization less then the 3-$\sigma$ level. 
This means that we did not find evidence for a statistically  significant 
deviation from the synchronization for any of these 15 objects.
The deviation from synchronization is statistically significant 
(at confidence level $>$99\%) only for the recurrent nova RS~Oph. 

For 22 S-type symbiotics we give clues as to what their orbital periods could be.
\end{abstract} 
 
 
\section{Introduction} 
The Symbiotic stars (SSs -- thought to comprise  a white dwarf                    
(WD) accreting from a  cool giant or Mira) represent the  
extremum of the interacting binary star classification. They                   
offer a 
laboratory in which to study                    
such important processes  as (i) mass loss from cool giants 
and the formation of Planetary Nebulae; (ii) accretion onto                    
compact objects, (iii) photoionisation  and radiative transfer                    
in gaseous nebulae, and (iv) nonrelativistic jets and bipolar 
outflows (e.g. Kenyon 1986; Corradi et al. 2003). 
                   
		    
On the basis of their IR properties, SSs have been classified into stellar continuum (S) 
and dusty (D or D') types (Allen 1982). The D--type systems contain Mira variables as mass donors.  
The D'--type  are  
characterized by an earlier spectral type (F-K) of the cool component
and lower  dust temperatures.  All mass donors in D'--type systems 
appeared to be very fast rotators (see Paper~I, Zamanov et al. 2006).
 
Our aims here are: 
{\bf (1)} to measure the projected rotational velocities (\vsi)  
and  the rotational periods 
(P$_{rot}$) of the  giants in a number of southern S-type SSs,  
using a cross correlation function (CCF) approach;  
{\bf (2)} to check whether their rotation is synchronized with the orbital period;
{\bf (3)} to provide pointers to the determination of binary periods  
(assuming co-rotation).  

This is the second in a series of papers exploring the rotation velocities  
of the mass donating (cool) components of SSs.                           
		                                    
\section{ Observations}
                   
We have observed 30 objects 
from the Belczy{\' n}ski et al.(2000) SS catalogue and have observed all S-type  
SSs  from the catalogue with 
$12^h<RA<24^h$, declination $<2^0$, and catalogue magnitude brighter than $V<12.5$.

The observations have been performed  with FEROS at the 2.2m 
telescope (ESO, La Silla).   
FEROS is a fibre-fed echelle spectrograph, providing a high resolution of 
$\lambda/\Delta \lambda=$48000, 
a wide wavelength coverage from about 4000~\AA\  to 8900~\AA\  in one exposure 
and a high throughput (Kaufer et al. 1999). 
The 39 orders of the echelle spectrum are registered with a 2k$\times$4k EEV CCD 

The present data have been collected from April to September 2004.
Table \ref{tab_log} gives a log of the observations. 
All spectra are reduced using the dedicated FEROS data reduction software 
implemented in the ESO-MIDAS system 
(www.ls.eso.org/lasilla/sciops/2p2/E2p2M/FEROS/DRS/). 
A few examples of our spectra are given in the Appendix (Fig.\ref{figEx}).

\begin{table}
\caption{Journal of observations. In the table are given as follows: 
 the name of the object, date of observation (YYYY-MM-DD), 
 the modified Julian Date (JD - 2400000.5) of the start  of the observation, 
 signal to noise ratio (S/N) around $\lambda$ 8500 \AA . }  
\begin{tabular}{lllrrr}
\hline
 object         & date-obs    &   MJD-OBS   &  Exp   &  S/N \\
                &             &              &  [sec] &      \\
\hline
         AR~Pav        &  2004-06-04 &     53160.3737 & 600  &  40 \\
         AS~255        &  2004-04-12 &     53107.3670 & 1200 &  45 \\
         AS~276        &  2004-04-12 &     53107.3967 & 600  &  45 \\
         AS~289        &  2004-06-05 &     53161.3071 & 600  &  40 \\
         AS~316        &  2004-06-07 &     53163.3600 & 1200 &  70 \\
         AS~327        &  2004-09-21 &     53269.0508 & 600  &  45 \\
  BD-21$^{\circ}$3873  &  2004-04-14 &     53109.2903 & 600  &  80 \\
  CD-36$^{\circ}$8436  &  2004-04-11 &     53106.3374 & 600  &  40 \\
  CD-43$^{\circ}$14304 &  2004-08-30 &     53247.2435 & 1200 &  70 \\
         FG~Ser        &  2004-06-03 &     53159.2731 & 1200 &  50 \\
         HD~319167     &  2004-06-07 &     53163.3905 & 600  &  45 \\
         Hen~2-374     &  2004-06-08 &     53164.3099 & 1200 &  40 \\
         Hen~3-1213    &  2004-04-12 &     53107.3049 & 1200 &  75 \\	     
         Hen~3-1341    &  2004-04-12 &     53107.3354 & 1200 &  50 \\
         Hen~3-1674    &  2004-06-07 &     53163.3103 & 1200 &  35 \\
         Hen~3-1761    &  2004-06-04 &     53160.2095 & 600  &  40 \\
         Hen~3-863     &  2004-04-11 &     53106.3067 & 1200 &  55 \\
         MWC~960       &  2004-06-08 &     53164.2738 & 1200 &  70 \\
         PN Ap 1-9     &  2004-06-07 &     53163.1986 & 1200 &  45 \\
         RS~Oph        &  2004-04-11 &     53106.3849 & 600  &  50 \\
         RW~Hya        &  2004-04-12 &     53107.2285 & 600  &  80 \\
         SS73~129      &  2004-06-29 &     53185.0660 & 1200 &  65 \\
         SS73~141      &  2004-06-29 &     53185.1058 & 1200 &  50 \\
         V2506~Sgr     &  2004-06-07 &     53163.2393 & 1200 &  45 \\
         V2756~Sgr     &  2004-06-08 &     53164.3968 & 600  &  40 \\
         V2905~Sgr     &  2004-06-28 &     53184.0858 & 1200 &  90 \\
         V3804~Sgr     &  2004-06-27 &     53183.9951 & 1200 &  40 \\
         V4018~Sgr     &  2004-08-31 &     53248.1800 & 1200 &  50 \\
         V4074~Sgr     &  2004-06-07 &     53163.2737 & 600  &  50 \\
         V919~Sgr      &  2004-06-03 &     53159.3108 & 1200 &  50 \\
                        &          &          &    \\           
\hline
  \label{tab_log}                                                      
  \end{tabular}                                                      
 \end{table}	  

\section{\vsi\ measurement techniques}

\begin{table*}
\caption{Rotational velocities of the red  giants in symbiotic  stars (measured in this paper). 
The spectral types  are from the catalogue  of
Belczy{\' n}ski et al.(2000), (B-V)$_0$   is the intrinsic colour (from Schmidt-Kaler 1982) 
for the corresponding  spectral type adopting luminosity  class  III in all cases, 
R$_g$ is the adopted radius of the giant [if not otherwise indicated, this is the average radius
for the corresponding spectral type taken  from Table~7  of van  Belle et al.(1999)].
P$_{orb}$ is the orbital period (see Sect.\ref{sec_ind}). \vsi (FWHM) is the projected rotational 
velocity of the cool giant as measured with FWHM method (see Sect.3.2).
$\sigma_{obs}$ is the observed width of the CCF.
\vsi\ is our measurement based on CCF method (see Sect.3.1), 
if other measurements of \vsi\ exist, they are also given
in the next column.
For objects with unknown orbit we give 
an estimation of the upper limit of the orbital period P$_{ul}$
(i.e. we expect P$_{orb}\lsim$P$_{ul}$, see Sect.\ref{clues}).
The upper part of this table contains objects observed with FEROS, the 
lower part - \vsi\ values are taken from the literature. 
}  
  \begin{tabular}{@{}lllrcrrrccr@{}}
\hline
\hline
 object        &Cool  & (B-V)$_0$&R$_g$     & P$_{orb}$&  \vsi &$\sigma_{obs}$ &  \vsi\   & other   & P$_{ul}$ & \\
               &      &          &          &          &  FWHM & CCF	       &  CCF	  &		&	   & \\
               &Sp.   &          &R$_{\sun}$& [days]   & \kms\ &[\kms]	 & [\kms ]  & [\kms ]	  &  [days]  & \\

\hline
\hline 
AR Pav         &    M5        &  1.59& 139.6   &  604.5 &   8.8\e2     &  6.208 &  8.1\e 1.5 &11\e2$^e$ 	  &   --- &  \\    
AS 255         &    K3        &  1.26& 20.5    &	&   9.7\e1     &  6.417 &  8.7\e 1.5 &  		  &   119 &  \\
AS 276         &    M4.5      &  1.60& 123.0   &	&   7.7\e2     &  6.382 &  8.6\e 1.5 &  		  &   724 &  \\
AS 289         &    M3.5      &  1.62& 89.0    &  451	&   9.5\e1     &  8.400 & 13.5\e 1.5 & 5.7\e1$^k$	  &   --- &  \\
AS 316         &    M4        &  1.62& 105.5   &	&   9.6\e1     &  6.843 &  9.8\e 1.5 &  		  &   545 &  \\
AS 327         &    M3        &  1.62& 71.5    &	&   7.7\e1     &  5.847 &  7.1\e 1.5 &  		  &   510 &  \\
BD-21$^{\circ}$3873   & K2   &  1.16& 20.8    &  281.6  &   6.6\e1     &  5.352 &  4.6\e 1.2 &5.4\e0.7$^f$	 &   --- &   \\
CD-36$^{\circ}$8436  &  M5.5 &  1.58& 144.0     &         &   6.6\e1.5   &  6.410 &  8.7\e 1.5 &  	 &   840 &   \\
CD-43$^{\circ}$14304 &   K5  &  1.51& 38.8    &   1448  &   7.2\e1     &  5.869 &  7.2\e 1.5 & $<$3$^g$ 	 &   --- &   \\ 								   
FG Ser         &    M5        &  1.59& 139.6   &   650  &   9.8\e1     &  6.756 &  9.6\e 1.5 & 8\e1$^h$, 7\e1$^k$ &  --- &  \\    
HD 319167      &    M3        &  1.62& 71.5    &	&   7.4\e2     &  6.041 &  7.7\e 1.5 &  	  &   472 &   \\
Hen 2-374      &    M5.5      &  1.59& 144.0   &	&   8.4\e1.5   &  5.712 &  6.7\e 1.5 &  	  &  1091 &   \\
Hen 3-1213     &    M2/K4     &  1.54& 50:?    &	&   9.1\e1.5   &  7.236 & 10.8\e 1.5 &  	  &   235 &   \\
Hen 3-1341     &    M2        &  1.61& 57.8    &	&   7.5\e1.5   &  6.423 &  8.7\e 1.5 &  	  &   336 &   \\
Hen 3-1674     &    M5        &  1.59& 139.6   &	&  56.0\e5     &  ..... & 52.0\e 5.2 &  	  &   135 &   \\
Hen 3-1761     &    M4        &  1.62& 105.5   &	&   9.3\e2     &  6.475 &  8.8\e 1.5 &  	  &   603 &   \\
Hen 3-863      &    K4        &  1.43& 45.0    &	&   7.9\e1     &  5.811 &  7.0 \e 1.5 & 		  &   326 &   \\		      
MWC 960        &    K9        &  1.55& 39.0$^a$  &	&   9.0\e1     &  6.436 &  8.7\e 1.5 &  	  &   225 &   \\
PN Ap 1-9      &    K4        &  1.43& 45.0    &	&   8.2\e1     &  5.792 &  6.9\e 1.5 &  	  &   329 &   \\
RS Oph         &    M0        &  1.56& 59.1    &  455.7 &  13.8\e1.5   &  7.624 & 11.7\e 1.5  & 		  &   --- &   \\
RW Hya         &    M2        &  1.61& 57.8    &  370.2 &  6.2\e1      &  5.837 &  7.1\e 1.5 &5.0\e1$^k$	  &   --- &   \\
SS73 129       &    M0        &  1.56& 59.1    &	&  8.9\e1      &  6.162 &  8.0\e 1.5 &  	  &   374 &   \\
SS73 141       &    M5        &  1.59& 139.6   &	&  7.9\e1      &  6.064 &  7.7\e 1.5 &  	  &   914 &   \\								       
V2506 Sgr      &    M5.5      &  1.59& 144.0   &	&  7.8\e1      &  6.284 &  8.3\e 1.5 &  	  &   874 &   \\
V2756 Sgr      &    M3        &  1.62& 71.5    & 243:?  &  4.2\e1.5    &  4.942 &  3.9\e 1.5 &  	  &   932 &   \\
V2905 Sgr      &    M5        &  1.59& 139.6   &	&  6.6\e1      &  5.708 &  6.7\e 1.5 &  	  &  1059 &   \\
V3804 Sgr      &    M5        &  1.59& 139.6   &	&  9.0\e3      &  ..... &  .....     &  	  &   ... &   \\
V4018 Sgr      &    M4        &  1.62& 105.5   &	&  7.2\e1.5    &  5.345 &  5.5\e 1.5 &  	  &   974 &   \\
V4074 Sgr      &    M4        &  1.62& 105.5   &	&  4.2\e1.5    &  4.871 &  3.5\e 1.5 &  	  &  1508 &   \\
V919 Sgr       &    M2        &  1.61& 57.8    &	&  7.4\e1      &  6.560 &  9.1\e 1.5 &  	  &   322 &   \\
    &           \\
\hline 
 SY  Mus        &   M5        &  &   139.6     &   624.5   & &  &  &   7\e1$^p$    &  & \\
 AG~Dra	        &   K2Ib or II&  & 30-40$^b$   &   554     & &  &  &  3.6\e1$^k$, 5.9\e1$^q$ &  & \\
 BX Mon         &   M5III     &  &   139.6     &   1401    & &  &  &  6.8\e1$^k$   &  & \\
 TX CVn         &   K5III     &  &   38.8      &   199     & &  &  &  8.9\e1$^k$   &  & \\
 T CrB	        &   M4IIIelips.& &  66\e11$^c$ &   227.57  & &  &  &  5.4\e1$^k$   &  & \\
 V443 Her       &   M5.5III   &  &   144       &   594     & &  &  &  4.5\e1$^k$   &  & \\
 CI Cyg         &   M5II or M5.5 &  & 189-236$^d$ & 854.5  & &  &  & 10.4\e1$^k$   &  & \\
 AG Peg         &   M3III     &  &   71.5      &   817.4   & &  &  &  4.5\e1$^k$   &  & \\
 V1329 Cyg      &   M6        &  &  147.9      &  956.5    & &  &  &  7.0\e2$^k$   &  & \\
 BF Cyg         &   M5III     &  &  75-280$^r$  &  756.8   & &  &  &  4.5\e2$^k$   &  & \\
                &  	      &  &	       &	   & &  &  &		   &  & \\
& & & & & & \\
\end{tabular} 
 \label{tab_vsi} 

$^a$-from  the fit in van Belle et al.(1999); 
$^b$Tomov et al.(2000);
$^c$Belczynski \& Mikolajewska(1998)
$^d$Kenyon et al.(1995);
$^e$Schild et al.(2001);
$^f$Smith et al.(1997); 
$^g$Schmid et al.(1998);
$^h$M{\"u}rset et al.(2000);
$^k$Fekel et al.(2003);  
$^p$Schmutz et al.(1994); 
$^q$de Medeiros \& Mayor (1999).
$^r$see Sect.\ref{sec_not}
 \end{table*}	  

\subsection{CCF method}
The projected rotational velocities have been derived
by cross-correlating the observed spectra with 
K0  numerical masks yielding a
cross-correlation function  whose width ($\sigma_{obs}$) is related to
broadening mechanisms such as stellar rotation and turbulence.

The emission lines do have an effect in the CCF and they must be  cleaned. 
They were cut off by fitting a continuum and replacing 
the  emission lines by the value of the fit. 
Note that the exact location  of the continuum is not important. 

The numerical K0-mask was constructed from a K0III synthetic spectrum 
in the region between $\lambda \lambda$ 5000-7000 \AA\
following the procedure described in  Baranne et al. (1979)
In the CORAVEL-type cross-correlation a binary mask is used  
as template instead a real spectrum. 
This binary (or CORAVEL-type) mask has been used in  many different 
data reduction software  for cross-correlation  
(ELODIE, CORALIE, and now HARPS). 
The K0-mask CCFs for the SSs observed 
here are plotted in the Appendix (Fig.\ref{CCF}) and $\sigma_{obs}$ are 
given in Table~\ref{tab_vsi}.

In order to use the observed width of the CCF ($\sigma_{obs}$) as 
an estimate of \vsi\ one needs to subtract  
the amount of broadening contributing to $\sigma_{obs}$ 
unrelated to the stellar rotation 
(convection, instrumental profile, etc), i.e. $\sigma_0$.
For FEROS spectra (see Melo et al. 2001, 2003): 

\begin{equation}
  v\: \sin i  = 1.9\, \sqrt{\sigma_{obs}^2 - \sigma_{0}^2} $\, \, {\rm \kms}$
\end{equation}

More details of the cross-correlation procedure are given in Melo et al. (2001), 
and also $\sigma_0$ is calibrated as a function of the 
$(B-V)$ for FEROS spectra and for stars with $0.6 < (B-V) < 1.2$.
For giants with $(B-V)<1.2$ in Table~\ref{tab_vsi} as well as in Paper~I, 
the Melo et al.(2001) calibration  has been adopted.  
However, the stars in our sample have  $(B-V)$ around 1.5 which is  beyond the
range of the calibration of Melo et al.(2001). 
For (B-V)$>1.2$, we will adopt constant $\sigma_0 =4.5$~\kms. 

This value has been adopted as a result of:
{\bf (i)} CCF measurements of few objects with known \vsi; 
{\bf (ii)} bearing in mind the results of Delfosse et al. (1998).
Using a similar template and 
method but lower resolution (which increases $\sigma_0$), 
Delfosse et al. (1998) have shown that $\sigma_0$ does vary over a range 
of $0.8 < (R-I) < 1.5$ (which corresponds to a spectral 
type from $\sim$M0 to $\sim$M6), decreasing from $\sim5.1$~\kms\ to 4.7~\kms; 
{\bf (iii)} bearing in mind the calibration of CCF and FEROS spectra
undertaken in Melo et al. (2001).

The errors on CCF \vsi\ measurements are dominated by systematic
effects rather than by photon noise. The error on \vsi\ comes from two main sources: 
uncertainties on the values of $\sigma_{obs}$  and $\sigma_0$. 
According to error estimate carried out by Melo et al.(2001), stars with $(B-V)<1.2$
have an error on \vsi\  less than 1.2 \kms.
For stars with $(B-V)>1.2$ we adopt a constant $\sigma_0$, 
which leads to an increase in our errors.
For such stars a conservative error of 1.5 \kms\ is assigned. 
For objects with \vsi$\ge$15 \kms\  the error on \vsi\ is $\pm$10\%.

In the case of the rapidly rotating Hen~3-1674 
the CCF rotation was extracted by a slightly different procedure as described for 
the fast rotators in Paper~I.
For V3804~Sgr we did not get a meaningful CCF due to the numerous emission lines
in the spectrum at the time of our observations. 
From measurements of the FeI 8689 line we obtain a rough estimate of 
FWHM(FeI 8689)$\approx0.58\pm0.15\!$ \AA\  
similar to the width of this line in AR~Pav and FG~Ser 
and corresponding to \vsi$\approx9\pm3$ \kms.

\subsection{FWHM method} 

Besides the cross-correlation procedure another method using 
full width at half maximum (FWHM) of spectral lines of observed and synthetic spectra 
was also applied. This procedure is similar to that described in Fekel (1997).
The spectra of 
K5III (T$_{eff}$=3950 K, $\log g=1.5$), 
M0III (T$_{eff}$=3985 K, $\log g=1.2$),
and M5III (T$_{eff}$=3424 K, $\log g=0.5$),
stars have been synthesized by using the code SYNSPEC 
(Hubeny, Lanz \& Jeffery 1994) in the spectral region  $\lambda$ 8750-8850 \AA.

LTE model atmospheres were extracted from Kurucz's grid (1993). 
The VALD atomic line database (Kupka et al. 1999) was used to create a line list 
for spectrum synthesis. The value of 3 \kms\  was adopted for the microturbulent velocity.   
A grid of synthetic spectra for projected rotational velocities from 0 \kms\ to 60 \kms\ was 
calculated. The FWHM of a dozen observed spectral lines 
has been measured and compared to the FWHM of spectral lines  
from the synthetic spectra. 
The results are given in the sixth column of Table \ref{tab_vsi}. 

\begin{figure}
 \mbox{}  
 \vspace{9.0cm}  
  \includegraphics{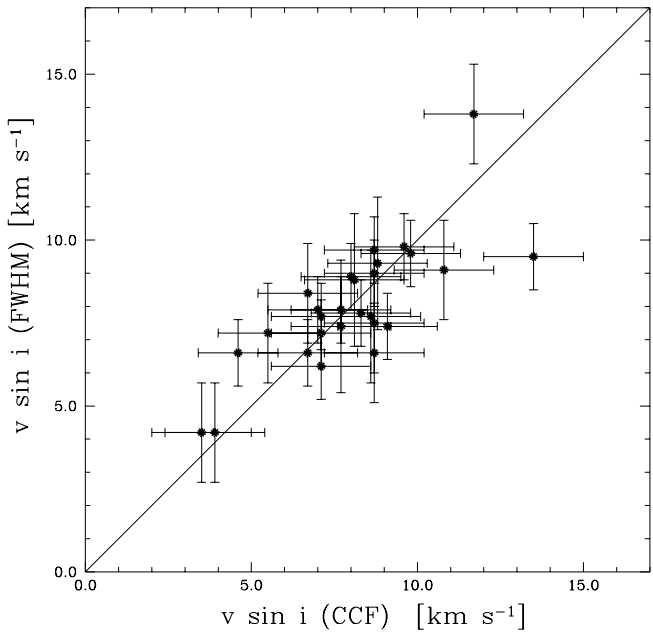}   
  \caption[]{Comparison between the CCF and FWHM measurements of \vsi.
  The straight line indicates \vsi(CCF)$=$\vsi(FWHM).
  }		    
\label{figC}     
\end{figure}

The comparison between the CCF and FHWM methods 
(see Fig.\ref{figC}) shows that the 
\vsi\ measurements agree well, with typical difference $\pm 1-2$ \kms,
as expected from the measurement errors.
The only exception is AS 289  (see also Sect. 6.1).

Because the CCF method uses many more lines  and a better defined 
mathematical procedure, in the  analysis
we will use \vsi\ derived with the CCF method. 

\section{Parameters of symbiotic stars}

\subsection{ Synchronization in SSs}

The physics of tidal synchronization for stars with convective
envelopes has been analyzed several times (e.g. Zahn 1977, and see the
discussion in Chapter 8 of Tassoul 2000). There are some
differences in the analysis of different authors, leading to varying
synchronization timescales. Here, we use the estimate from Zahn (1977,
1989). The synchronization timescale in terms of the period is
\begin{equation}
\tau_{syn} \approx 800 \left( \frac{ M_g  R_g}{ L_g}\right)^{1/3} 
\frac{M_g^2 (\frac{M_g}{M_2} + 1)^2}{R_g^6} P_{orb}^4\ \; \; \; \; {\rm years}
\end{equation}
where M$_g$ and M$_2$ are the masses of the giant and white dwarf
respectively in Solar units, and R$_g$ and L$_g$ are the radius and
luminosity of the giant, also in Solar units. The orbital period P$_{orb}$ is
measured in days.

In Table~\ref{tab_syn} we calculate this timescale for a few  
representative cases. We point out that these values serve as an estimation.
From the $R_g^6$ dependence of $\tau_{syn}$ in Eq. 1, one can expect large uncertainties in
the synchronization timescale as a result of the uncertainties in the red
giant radius. There is also a $P_{orb}^4$ dependence, which can introduce
large uncertainties, but  $P_{orb}$ is much more accurately known
than $R_g$.

As can be seen in Table~\ref{tab_syn}, for S--type 
SSs depending on the binary parameters, $\tau_{syn}$ can be as short as $<10^4$ yr and longer than
10$^7$ years. 
Short $\tau_{\rm syn}$ can be expected 
when the  radius of the mass donor
is a significant fraction of the orbital separation. 

The lifetime of the symbiotic phase for a red giant or AGB star is 
around $10^5$year (see e.g. Yungleson et al. 1995), 
which is comparable with $\tau_{\rm syn}$.

\begin{table}  
\caption{The synchronization time, $\tau_{syn}$, calculated following  
Eq.1 for a few representative cases. The typical values for 
L$_g$ and  R$_g$ of K5~III and M4III stars are adopted following
van Belle et al.(1999). 
 }  
 \begin{tabular}{@{}lrrrrrr@{}}  
\hline
\hline
Mass donor & L$_g$         & M$_g$     & R$_g$  & P$_{orb}$  &  M$_{2}$  & $\tau_{syn}$  \\
Sp. type & [L$_{\odot}$] &[M$_\odot$]& [R$_\odot$] &  & [M$_\odot$]  &   [yr]    \\
\hline
K5~III & 350  & 2 & 40 &  200 d & 1.0 & 7000     \\
       &      &   &    & 1500 d & 1.0 & 2.10$^7$ \\       
       &                                         \\
M4~III & 1380 & 1 & 105 &  500 d & 0.5 & 140     \\ 
       &      & 2 &     &  500 d & 1.0 & 700     \\
       &      & 3 &     &  500 d & 1.0 & 3200    \\
       &      & 2 &     & 1500 d & 1.0 & 60000   \\
 & & \\
 \hline 						 
 \label{tab_syn} 
  \end{tabular}									   				      
  \end{table}



\subsection{Inclination}
\label{sec_incl}
In our calculations, we assume that the rotational axis of the red giant 
is perpendicular to the orbital plane (see Fekel 1981; Hale 1994; Stawikowski 1994).
To calculate the rotational periods of the mass donors (P$_{rot}$)
we need to know the inclination of the orbit ($i$) and R$_g$.

For eclipsing binaries we can adopt inclination $i=70^0 - 110^0$, 
which produces small errors in  $\sin i$. For such cases 
we will assume that  $\sin i$  is in the range 0.94-1.00.
For other cases we use results from spectropolarimetric observations
or radial velocity measurements.  

\subsection{Radius of the cool component}
\label{sec_radii}

For a few objects, the radii of the cool components are derived 
from model calculations. 
If the radius of the giant (R$_g$) is not known, we will use 
the average radius for the corresponding  
spectral type taken  from van  Belle et al.(1999),  
always adopting luminosity  class III and error $\pm5\%$.

For the M giants these values are similar to the stellar radii of M giants 
in the Hipparcos catalogue as calculated by Dumm \& Schild (1998).
Dumm \& Schild (1998) have also shown that the radius of the
M giants depends on the mass. 
For  9 objects included in our sample, the  masses of the M giants are listed
in Miko{\l}ajewska(2003). In Table~\ref{tab_radii}, 
we  compare the values adopted here 
(following van  Belle et al. 1999) with
the results of Dumm \& Schild (1998).

\begin{table}  
\caption{Stellar radii of M giants.
In column 3 are given R$_g$ (following van  Belle et al. 1999),
in column 4 - mass of the M giants (from Miko{\l}ajewska 2003),
in column 5 -  the range of R$_g$ for the corresponding mass and spectral
type (following Dumm \& Schild 1998). }
 \begin{tabular}{lrrccrr}  
\hline
Object & Mass donor & R$_g$         & M$_g$     & R$_g$       &  \\
       &            & [R$_{\odot}$] &[M$_\odot$]& [R$_\odot$] &  \\
1      & 2          & 3             & 4         & 5           &  \\
\hline
FG~Ser &  M5III & 140 &    1.7     & 110-150 & \\
AR~Pav &  M5III & 140 &    2.5     &  90-240 & \\
AG~Peg &  M3III &  72 & $>$1.8     &  60-90  & \\
BX~Mon &  M5III & 140 & 3-3.7      &  150-200  & \\
SY~Mus &  M5III & 140 & 1.3        &  60-100   & \\
RW~Hya &  M2III &  58 & 1.6        &  40-80    & \\
AR~Pav &  M5III & 140 & 2.5-3      &  130-160  & \\
BF~Cyg &  M5III & 140 &  1.8       &  110-140  & \\
AG~Peg &  M3III &  72 & $\gsim$1.8 & 60-90    & \\
 \hline 						 
 \label{tab_radii} 
 \end{tabular}									   				      
 \end{table}

As visible in most cases, the adopted radii (following van  Belle et al. 1999)  
are in agreement with the mass-dependent radii of Dumm \& Schild (1998). 
There are 2 cases 
where  the adopted values are outside of this range:
BX~Mon and SY~Mus (see Sect.\ref{sec_ind}).

\section{Results}

Our measurements of \vsi\ together with data collected from the literature,
are summarized in Table~\ref{tab_vsi}. 
Our sample thus contains 39 objects (29 with \vsi\ measured by us and 10 
taken from the literature). This sample should have no biases in 
the rotational speed of the cool giant, even though the sample is flux limited. 


In our sample of S-type SSs, the projected  rotational velocities 
of  the mass donors are from $3.5$~\kms\ up to 52~\kms.  
Including our data and the data from the literature, we obtain
for \vsi: 
mean$=\!8.7$ \kms, median$=\!7.7$ \kms, standard deviation$=\!7.4$ \kms.

If we exclude the 2 slowest and the 2 fastest rotators we get: mean \vsi$=7.6\pm1.8$ \kms. 
In effect, 90\% of the mass donors in S-type SSs with measured rotation 
have \vsi\ in the interval  $4.5\le$\vsi$\le11.7$ \kms. 
For the southern S-type symbiotics (29 objects, flux limited sample, FEROS observations)
the values are similar: median \vsi=8.1 \kms, 
90\%  of the objects in the interval  $4\le$\vsi$\le11.7$ \kms.

\section{S--type symbiotics with known orbital periods}
\label{sec_ind}

In our sample, there are 17 objects with known orbital periods 
and 1 where this is inferred (V2756 Sgr). 
In this section we compare the orbital
periods with the rotational periods of the mass donors  object by object.
The upper-lower limits of P$_{rot}$ are calculated from: 
\begin{equation}
 P_{rot}=\frac { 2 \pi (R_g\pm e_1) (\sin i \pm e_2) }{v\: \sin i \mp e_3},
\end{equation}
where $e_1$, $e_2$, and $e_3$ are the corresponding errors in 
R$_g$, $\sin i$, and \vsi\ respectively. 

In the statistical analysis (Sect.\ref{Sec_sta}), 
for  P$_{rot}$  we will use the average value between these upper-lower limits 
and will consider them as corresponding to $\pm 1 \sigma$ error. 



 \begin{figure*}
 \mbox{}  
 \vspace{15.0cm}  
  \includegraphics{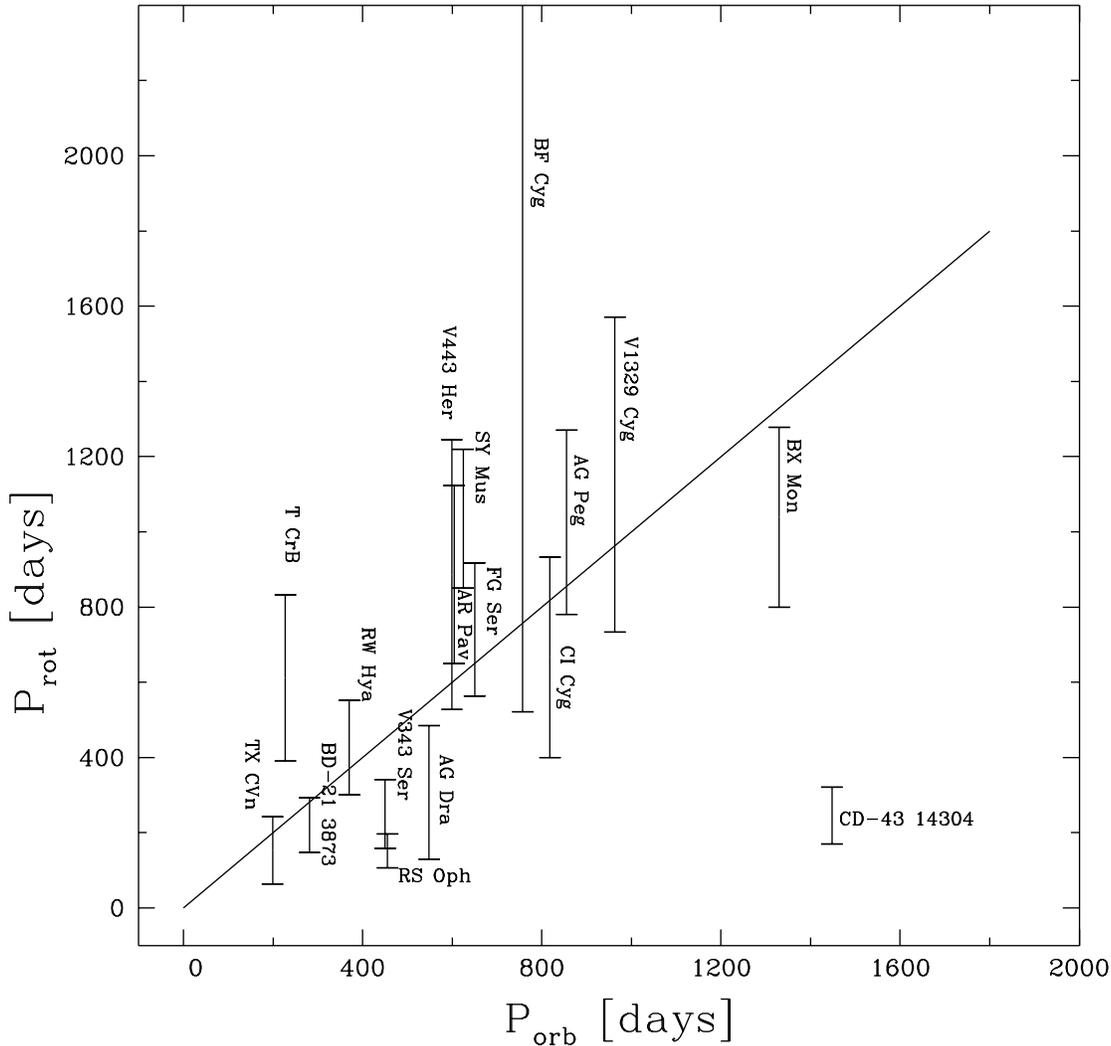}   
  \caption[]{The rotational period of the red giant (P$_{rot}$) 
  versus the orbital period (P$_{orb}$) of the 
  17 objects in our sample.  The solid line corresponds to 
  P$_{rot} =$P$_{orb}$.  Most objects are close to this line,
  which indicates that they are synchronized. There are 2 objects
  which deviate considerably from that rule (RS~Oph and CD-43$^\circ$14304).
  }		    
\label{P-P}     
\end{figure*}	      

\subsection{Objects with known orbital periods observed by us}
\label{sec_our}

{\bf AR Pav} 
is an eclipsing binary with P$_{orb}\approx$ 604.5 days (Bruch et al. 1994).
The giant's radius derived from the eclipse is
R$_{g} =137\pm 20\;R_{\sun}$ (Quiroga et al. 2002), which is very close
to the value adopted in Table~\ref{tab_vsi} from the spectral type of the cool giant
and to R$_g=139\pm10\;R_{\sun}$ (Skopal 2005).
We calculate 649 $< $P$_{rot}<$ 1120 days.

{\bf AS 289 } (V343 Ser)    
has  orbital period P$_{orb}=450.5\pm2.2$~d, 
$e=0.135\pm0.046$,
and orbital inclination of $17^\circ-23^\circ$ (Fekel et al. 2001). 
Kenyon \& Fernandez-Castro (1987) gave spectral type M3.9III, 
but more recently M3.5 is assigned from M{\" u}rset \& Schmid (1999).
Assuming it is a normal M3.5III star, and using  
our CCF value for \vsi, we obtain $159<$ P$_{rot} < 341$ days. 

In a binary with  an eccentric orbit
the synchronization is reached at  a value of
P$_{rot}<$P$_{orb}$ (pseudosynchronization, see Hut 1981).
Following Eq.~42 and Eq.~43 of Hut (1981), in
AS 289 the pseudosynchronization is expected
at P$_{rot}=0.90\:$P$_{orb}$.

It is worth noting that  Fekel et al.(2003) 
give a value of \vsi$= 5.7 \pm 1$ \kms\ and our FWHM method give
\vsi$=9.5\pm 1$ \kms.  
An independent measurement of \vsi\ with better spectroscopic
resolution will be valuable.

{\bf  BD-21$^{\circ}$3873} has P$_{orb}=281.6\pm1.2$  days, 
sin $i=0.87$ (Smith et al. 1997). 
We derive  $147 < $ P$_{rot} < 293$ days. The object is thus synchronized 
within the measurement errors.

{\bf RW~Hya}  is an eclipsing binary with P$_{orb}$ = 370.4\e0.8 days (Schild et al. 1996). 
Supposing  $i=70^0-90^0$, we calculate that P$_{rot}$ is in the interval 
301 - 552 days. The object is thus synchronized 
within the measurement errors.

{\bf  FG Ser} (AS~296) is an eclipsing binary with P$_{orb} = 650\pm 5$ days 
(M{\" u}rset et al. 2000). Supposing  $i=70^0-90^0$, 
we calculate P$_{rot} \approx 563 - 918$ days.  
The object is thus synchronized within the measurement errors.

\subsection{Objects with known periods, not observed by us}  
\label{sec_not}
Fekel, Hinkle \& Joyce (2004) published values of \vsi\ for 13 S-type SSs. 
In this section, if not otherwise stated, we use their measurements. 

{\bf SY Mus} is an eclipsing binary with  $i=95^0.8\pm1.7^0$ 
(Harries \& Howarth 1996, 2000),
P$_{orb}$ = 624.5 days and  \vsi\ = 7\e 1 \kms\ 
(Pereira et al.1995; Schmutz et al. 1994; Kenyon \& Mikolajewska 1995). 
We calculate P$_{rot} \approx 830-1230$ days.
[If we use  R$_g\approx 80$~R$_{\odot}$ (see Sect.\ref{sec_radii}), 
we get P$_{rot}\approx  480-705$ days, a range which includes
the orbital period.]

{\bf AG Dra} is a yellow SS with spectroscopic P$_{orb} = $ 548.5\e2~d 
(Friedjung et al. 2003; Fekel et al. 2000b),
and  $i\approx 30^0-45^0$ (Mikolajewska et al.1995).
The cool component is a K2~Ib~or~II (Zhu et al. 1999).
There are two measurements of \vsi : 
5.9\e1.0 \kms\  (de Medeiros \& Mayor 1999) and 3.6\e1.0 \kms\ (Fekel et al. 2004).
We adopt  \vsi$ = 4.8\pm1.0$ \kms.  
Skopal (2005) calculated R$_g = 33\pm11~R_{\sun}$. 
Following  Tomov et al.(2000) and the references therein  
we can adopt  R$_g = 30-40~$R$_{\sun}$. 
We then estimate  P$_{rot} \approx  130 - 485$ days.

{\bf  TX CVn  } has P$_{orb}=199\pm3$~d, $e=0.16\pm0.06$, 
with the cool component a normal  K5III star, 
and inclination $20^\circ<i<70^\circ$ (Kenyon \& Garcia 1989).
The luminosity class could also be II or Ib (Zhu et al. 1999).
We calculate P$_{rot} \approx 64-242$ days.
Pseudosynchronization (see Hut 1981)
is expected at P$_{rot}=0.87$P$_{orb}$.
The object is thus (pseudo)synchronized 
within the measurement errors.  

{\bf  AG~Peg} has  P$_{orb}=818.2\pm1.6$ days and
$e=0.110\pm0.039$ (Fekel et al. 2000a), normal M3III and 
$i\sim40^\circ-60^\circ$ (Kenyon et al. 1993).
We calculate 
P$_{rot}\approx  400-933$ days. The object is thus synchronized 
within the measurement errors.
  
{\bf  V1329 Cyg  }   has     P$_{orb}=963.1\pm9.8$ days (Chochol \& Wilson 2001)
and  inclination $i=86^0\pm 2^0$ (Schild \& Schmid 1997). 
We calculate P$_{rot}\approx  734 - 1571$ days. The object is thus synchronized 
within the measurement errors. 
  
{\bf  T CrB} 
has   P$_{orb}=$ 227.57 d (Fekel et al. 2000a),  
and R$_g=66\pm11\;R_\odot$ (Belczynski \& Mikolajewska 1998). 
[It should be noted that Skopal (2005) gives R$_g=75\pm12 (d/960$pc),
and that a normal M4III would have R$_g\approx105.5$R$_\odot$.]
The inclination of the system is $i\approx 65^0-70^0$ (Stanishev et al. 2004). 

We calculate  P$_{rot}\approx  391 - 832$ days, which differs
from the expectations for synchronization.
This result is unexpected, however it appears that this deviation is not statistically significant 
(see Table~\ref{T_p}). From Eq.1 we get $\tau_{syn}<100$ yr. 
The red giant in T~CrB is ellipsoidally shaped (Yudin \& Munari 1993).
This means that the object has to be synchronized.

{\bf  BX Mon }  The orbital parameters
of this eclipsing system are: \\
P$_{orb}=1401$ days, eccentricity $e=0.49$ (Dumm et al. 1998); \\
P$_{orb}=$1259~d, $e=0.44$ (Fekel et al. 2000a). \\

We calculate  P$_{rot}\approx 800 - 1278$ days
[if we use  R$_g\approx 170$~R$_{\odot}$ (see Sect.\ref{sec_radii}), 
we get P$_{rot}\approx 970 - 1550$ days].
BX~Mon is the only SS with 
considerable orbital eccentricity in our sample.
SSs with P$_{orb}>800$ days tends to eccentric orbits (Fekel et al. 2007).
In a binary with  $e=0.44$, 
the pseudosynchronization is expected at  about
P$_{rot}=0.46\;$P$_{orb}$ (see Hut 1981).

The time scale for 
synchronization in SSs is $\sim$10 times shorter then the
circularization time (see Schmutz et al. 1994).
All this implies  that in this system 
the red giant is more or less (pseudo)synchronized, but the orbit is  
not circularized yet. 
It has therefore to be in the process of circularization.

{\bf  V443 Her} is not eclipsing, with P$_{orb}=599.4\pm2.1$ days 
(Fekel et al. 2000b), 
viewed at an inclination $i\sim 30^{\circ}$ (Dobrzycka et al. 1993). 
We calculate  P$_{rot}=528-1245$ days.
The red giant is thus synchronized within the measurement errors. 

{\bf  CI Cyg }  is an eclipsing binary with orbital 
period 855.6 days and orbital separation $a=2.2$~au (Kenyon et al. 1995). 
Kenyon et al.(1995) have shown that the mass donor is
an M5II asymptotic branch giant, filling its tidal surface.
They  calculated from the eclipse that 
R$_g/a\sim 0.4-0.5$ which means that R$_g\approx 189-236$ R$_{\odot}$.
We calculate P$_{rot}=780-1270$ days. The object is thus synchronized 
within the measurement errors.

{\bf  BF~Cyg }  is an eclipsing binary with  P$_{orb}=757.3$ days 
and inclination $i\approx70-90^0$ (Pucinskas 1970, Skopal et al. 1997).    
If the cool component is a normal M5III giant we expect 
R$_g=139.6\pm5\%$ R$_{\odot}$. 
Mikolajewska et al.(1989) suggested  R$_g\sim75$ R$_{\odot}$,
however Skopal et al. (1997) give R$_g=260\pm20$ R$_{\odot}$.
Using \vsi$=4.5\pm2$ \kms, we calculate P$_{rot} \sim 560 - 5660$ days.

There are signs that the red giant is ellipsoidally shaped
(Yudin et al. 2005). $\tau_{syn}$ for such an object will
be short, from Eq.1 we get $\tau_{syn}<7000$~yr. 
It means that the red giant has to be tidally locked and most probably 
the R$_g$ is closer to a value $\sim75$ R$_{\odot}$.

\section{Questionable Objects}
\label{sec_q}

{\bf RS~Oph} has orbital period P$_{orb}=455.72\pm0.83$ days
(Fekel et al. 2000a). The inclination is about 30--40$^{\circ}$ 
(Dobrzycka \& Kenyon1994; Dobrzycka et al. 1996).
This will give a  $108<$P$_{rot}<197$ days,
i.e. 2--3 times less than the orbital period.

RS~Oph is a peculiar SS exhibiting different types of activity -- recurrent nova eruptions, 
jet or blob ejections, flickering 
(see Bode et al. 2006 and references therein). 

This is one of the two  objects in our sample whose 
deviation from synchronization is statistically significant 
(at confidence level $>$99\%, see Table~\ref{T_p}). 
The red giant in  
RS~Oph seems to rotate faster than the orbital period, 
which means that it has to be in a process of deceleration
(the expected $\tau_{syn}\le5.10^4$ yr).

While there are no doubts about P$_{orb}$, any of the other parameters 
(\vsi, inclination, red giant radius) have to be checked 
with independent measurements. 

It is noteworthy that our experiments to measure \vsi\ with CCF 
and spectra from different epochs showed that \vsi\ could even 
be  higher (up to 14.5\e1.5 \kms).

{\bf CD-43$^{\circ}$14304} 
Schmid et al.(1998) reported a circular orbit, P$_{orb} = 1448\pm100$ days.
They argued that the presence of phase-dependent H$\alpha$ variations is
attributable to occultation effects, suggesting
a relatively high inclination, $i > 45^0$. 
Spectropolarimetry   (Harries \& Howarth 2000) gives two possible values
for the inclination of the system $i=57^0\pm5^0$  or $i=122^0\pm48^0$. The second corresponds to an improbably 
large eccentricity for the orbit. 
Therefore  we can accept  $\sin i= 0.79 - 0.88$.
With our value of \vsi\ we calculate  P$_{rot} \approx 170 - 321$ days. 

However, Schmid et al.(1998) 
found no evidence for rotational broadening in the cool-giant spectrum 
and were able to place an upper limit on \vsi$<3$ \kms. 
We measured   \vsi$= 7.2 \pm 1.5$ \kms\ the CCF looks good 
(see Fig. \ref{CCF}) and FWHM method gives a similar result. 
If our CCF result is wrong and that of Schmid et al.(1998)
is a correct one, then the object is probably synchronized. To avoid 
confusion we will exclude this star from the analysis, 
but an independent check of \vsi\ would be valuable.

{\bf V2756 Sgr} (Hen 2-370, SS73 145). The parameters of the system 
are not well known.
We calculate P$_{rot} < 1595$ days. 
A photometric period of 243 days is supposed in Hoffleit (1970).
This photometric period is not confirmed with radial velocity measurements 
to be that of the orbit. If this is the orbital period 
and the red giant is tidally locked, then the inclination of the system 
would be about $i\sim 15^0$.  

{\bf Hen~3-1674} is the fastest rotator in our sample.
The catalogues indicate  that the 
mass donor is probably M5III star (see Table~\ref{tab_vsi} and the references therein).
A normal M5III star would have 
R$_g \approx 90-160$  \rsun\ (van Belle et al 1999) 
and M$_g \approx 1-3$~\msun, which 
means a break-up velocity of 30-60 \kms. If our measurements and the adopted 
parameters of the system are correct then Hen~3-1674 rotates close to 
its critical velocity and has rotation similar to that of 
D'-type SSs (see Paper~I).

\section{Are the mass donors in S--type symbiotics synchronized (co-rotating) ?}  
\label{Sec_sta}
Fig.~\ref{P-P} shows the rotational period versus the orbital period of the 17 objects in our sample, 
with a straight line indicating the co-rotation (i.e. P$_{rot}$=P$_{orb}$). 
Most objects are close to this line, which suggests that they are synchronized. 
In Table~\ref{T_p} are given the individual deviations as well as the corresponding probability 
that the deviation is random. 
9 objects are synchronized within the measurement errors (1-$\sigma$ level).
4 objects have deviations between 1 and  2-$\sigma$. Generally, 15 out of 17 are  
within the 3-$\sigma$ level. The two objects
that are outside of the 3-$\sigma$ level are RS~Oph and CD-43$^\circ$14304.

The standard $\chi^{2}$ test gives a probability of  p$(\chi^{2})<10^{-6}$ that 
the co-rotation straight line fits the data points when we use all 17 objects. 
Here we have considered the errors of the rotational period only, 
the errors of P$_{orb}$ are supposed to be small (usually they are $<$2\%). 
The straight line cannot be rejected as a fit to the data at  
more than 90\% confidence level, p($\chi^{2}$)$=$0.127,
when the two deviating objects 
(CD-43$^\circ$14304 and RS~Oph) are excluded.
The $\chi^{2}$  statistic tests how well the data are described 
by the model, assuming that all the deviations are due to measurement errors
and not to intrinsic scatter.

When intrinsic scatter is allowed, as we expect 
to be the more realistic situation, one can apply the weighted least squares estimator 
(e.g. Akritas \& Bershadi 1996) to find the slope ($\beta$) and 
the intercept ($\alpha$) of a linear fit in the form \\ 
P$_{rot}=\alpha+\beta$P$_{orb}$ to the data points with measurement 
errors and non-negligible intrinsic scatter. When all objects are included, 
we get $\alpha=204\pm49$ and $\beta=0.14\pm0.07$, which is not consistent 
with P$_{rot}$=P$_{orb}$ ($\alpha=0$, $\beta=1.0$). 
The situation changes if CD-43$^\circ$14304 
is removed from the sample, then $\alpha=-110\pm77$ and $\beta=0.93\pm0.16$. When 
the deviating recurrent nova RS~Oph is removed as well, we obtain 
$\alpha=-32\pm78$ and $\beta=1.01\pm0.17$. The last result is fully consistent with 
P$_{orb}$=P$_{rot}$ line ($\alpha=0$, $\beta=1$). 

Although the synchronization line seems to fit well the data, we still
cannot rule out the possibility that one or more outlying objects may
determine a false correlation between the variables. Therefore, a rank
correlation test appears appropriate. The Spearman rank correlation
coefficient between the two periods is 0.61 when all 17 objects were
included, and about 0.82 for 16 and 15 objects only. This result implies a
significant correlation between the variables, since the p-values are less
than 0.02 and 0.002 respectively. The similar Kendall-tau test gives
similar results, with p-values less than 0.01 for all cases.

The objects that deviate significantly from the P$_{orb}$=P$_{rot}$ line are
RS~Oph (a peculiar SS as noted above)  and 
CD-43$^\circ$14304 (with possible error in the \vsi\ measurement, see Sect.\ref{sec_q}). 
Taking into account the uncertainties of their P$_{rot}$, one finds that the probabilities for a 
random deviation are less than 0.01 for them 
(see Table~\ref{T_p}), while for all the other 15 stars this idea cannot be rejected 
at the 0.01 significance level. 
In other words the null hypothesis that all S-type SSs
with well measured \vsi\ are synchronized 
(excluding RS~Oph and CD-43$^\circ$14304) cannot be rejected at the 99\% confidence level.

\begin{figure}
 \mbox{} 
 \vspace{7.5cm} 
 \includegraphics{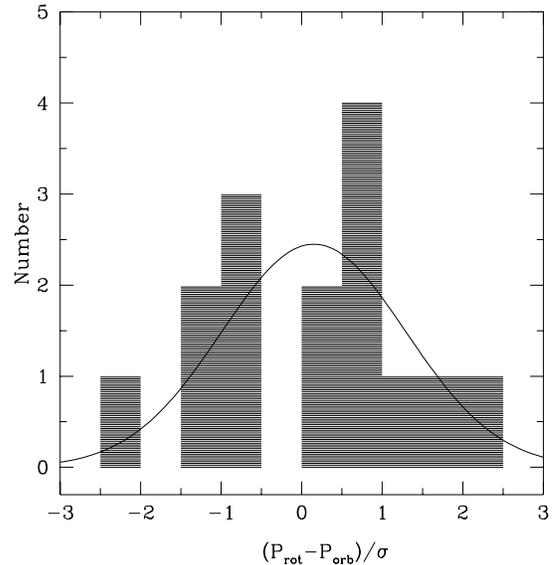} 
 \caption[]{The distribution of (P$_{rot}$-P$_{orb}$)$/\sigma$ for 15 symbiotic stars 
            and the fitted gaussian ($\sigma$ is the error of P$_{rot}$). The K-S statistic gives only a 38\% probability 
            that the parent distribution deviates from the normal one (not statistically significant).}
	\label{fig_distr}    
\end{figure}

An additional test that may give some clues about the extent of synchronization of the periods is 
the K-S test of how much (P$_{rot}-$P$_{orb}$)$/\sigma$ deviates from the normal distribution.
Even when all 17 stars are included, the p-value of the K-S statistics is slightly less than 10\%.
When only the above 15 objects are included (see Fig.\ref{fig_distr}), 
the distribution becomes much narrower (mean\,$=\!0.15$, $\sigma$=1.23,
K-S statistics is 0.38), with a standard deviation comparable to the 
measurement errors of P$_{rot}$, $<\sigma/P_{rot}>=0.53$.
This means that the hypothesis that the sample comes from a normal distribution cannot be 
rejected statistically.

We see that the null hypothesis for synchronization of the red giants in symbiotic 
stars cannot be rejected statistically (except probably for RS~Oph).


 \begin{table}
  \caption{The deviations of the individual objects from the line
   P$_{rot} = $P$_{orb}$. In the second column the deviations are given 
  in units of $\sigma$, where $\sigma$ is the individual error of P$_{rot}$. 
  The third column gives the probability for random deviation. 
   To reject the null hypothesis (synchronization) at 99\% confidence level,  
   p($\chi^{2}$) has to be smaller than 0.01.}
  \begin{tabular}{lrr}
  \hline
  \hline

  Object &  deviation & p($\chi^{2}$)  \\
         &   [$\sigma$]         &  \\
  \hline

SY Mus               & 2.2  & 0.026	   \\
AG Dra               & 1.4  & 0.175	   \\
BX Mon               & 1.2  & 0.223	   \\
TX CVn               & 0.5  & 0.605	   \\
T CrB                & 1.7  & 0.082	   \\
V443 Her             & 0.8  & 0.423	   \\
CI Cyg               & 0.7  & 0.489	   \\
AG Peg               & 0.6  & 0.569	   \\
V1329 Cyg            & 0.5  & 0.651	   \\
BF Cyg               & 0.9  & 0.361	   \\
AR Pav               & 1.2  & 0.233	   \\
V343 Ser             & 2.2  & 0.028	   \\
BD-21$^{\circ}$3873  & 0.8  & 0.399	   \\
CD-43$^{\circ}$14304 & 15.9 & $<\!10^{-6}$ \\
FG Ser               & 0.5  &  0.610	   \\
RS Oph               & 6.7  & $<\!10^{-6}$ \\
RW Hya               & 0.4  & 0.655	   \\

  \hline
  \end{tabular}
  \label{T_p}
 \end{table}





\section{Clues to the orbital periods}
\label{clues}
Up to now, out of 188 SSs, the orbital elements and binary periods  
are well known for  $\sim$40 objects only (and they are all S--type). 
The derived orbital periods are 
in the range 200 -- 2000 days (Miko{\l}ajewska 2003).

Because the orbital periods of the majority of SSs 
are unknown, an indirect method to obtain P$_{orb}$ is to measure  
\vsi. If the mass donors in SSs are co-rotating (P$_{rot}=$P$_{orb}$),  
we can find clues for the orbital periods via the simple relation  
P$_{orb}$ v$_{rot}$  = 2$\pi$R$_g$, where P$_{orb}$ is the orbital period, 
v$_{rot}$ and R$_g$ are the rotational velocity and radius of the giant, 
respectively. 
The underlying suppositions
are: {\bf (1)} corotation (see Sect.~\ref{Sec_sta})
and  {\bf (2)} rotational axis of the red giant 
perpendicular to the orbital plane (see Sect.~\ref{sec_incl} and references therein).

It could be useful in the case of eclipsing binaries, where  
$\sin i\approx1$ and \vsi$\approx$v$_{rot}$, however in such cases it is easy to find 
P$_{orb}$ with photometry. If the inclination is unknown, we can  
only put an upper limit,  P$_{orb} \lsim$P$_{ul}$, 
where P$_{ul}=2\pi$R$_g/$\vsi. To see how useful 
this predictor is, we calculated the upper limits for 
the objects with known orbital periods,
and we get P$_{ul}/$P$_{orb}\sim 0.5-1.5$. 

For the objects with unknown periods, the upper limits for P$_{orb}$ calculated 
in this way are given in  the last column of Table~\ref{tab_vsi}. 
These upper limits  are in the 
interval  100--1500 days. Most of them  are similar to those of 
the measured orbital periods in S-type SSs. 
However, it seems that for AS 255 and Hen~3-1674, 
P$_{orb}$ could be as short as $\lsim 150$ days.

\section{Conclusions}  
 
We have observed 30 S-type symbiotic stars 
with the FEROS spectrograph.
We have measured the rotational velocities of the mass donors for 
29 of them by the means of the CCF method 
(for V3804 Sgr we did not get a meaningful CCF). 
The results of the CCF method have been checked by the
FWHM measurements. The main results are:

(1) The projected, rotational velocities of the cool components in S-type
symbiotic stars are from $3.5$ \kms\ up to 52 \kms\  and 90\% of them 
are in the interval from 4.5 \kms\ to 11.7 \kms. 

(2) In our sample of 17 S--type SSs with known orbital periods, 
9 are synchronized within the measurement errors (1-$\sigma$ level).
If we exclude the doubtful object CD-43$^\circ$14304 and the 
recurrent nova RS Oph, 
{\bf all remaining 15 objects are synchronized within the 3-$\sigma$ level}. 
In other words, 
the null hypothesis that these 15 objects are synchronized 
can not be rejected statistically at the 99\% confidence level. 

(3) Among all objects with \vsi\ measured by us
the deviation from synchronization is statistically significant 
(at the 99\% confidence level) only for RS~Oph. The red giant in 
this peculiar object seems to rotate faster than the
orbital period.
 

(4) For 22 S-type SSs with unknown parameters, 
we give clues as to what their orbital periods could be. 
  
In future it will be interesting:
(i) to  measure the projected rotational
velocity of the cool giants in more symbiotic stars;
(ii) to compare their rotational velocity with that of the isolated giants and
those in other binary systems.

\section*{Acknowledgments}  
  
This research has made use of SIMBAD, IRAF, and  Starlink.  
RZ was supported by a PPARC Research Assistantship  
and MFB was a PPARC Senior Fellow.  AG acknowledges the receipt of 
Marie Curie Fellowship and Marie Curie European Re-integration Grant 
from the European Commission. This work was partially complete when Dr John Porter passed away in June 2005 and we dedicate this paper to his memory.

\section*{APPENDIX} 

\begin{figure*}
 \mbox{}  
 \vspace{18.0cm}  
  \includegraphics{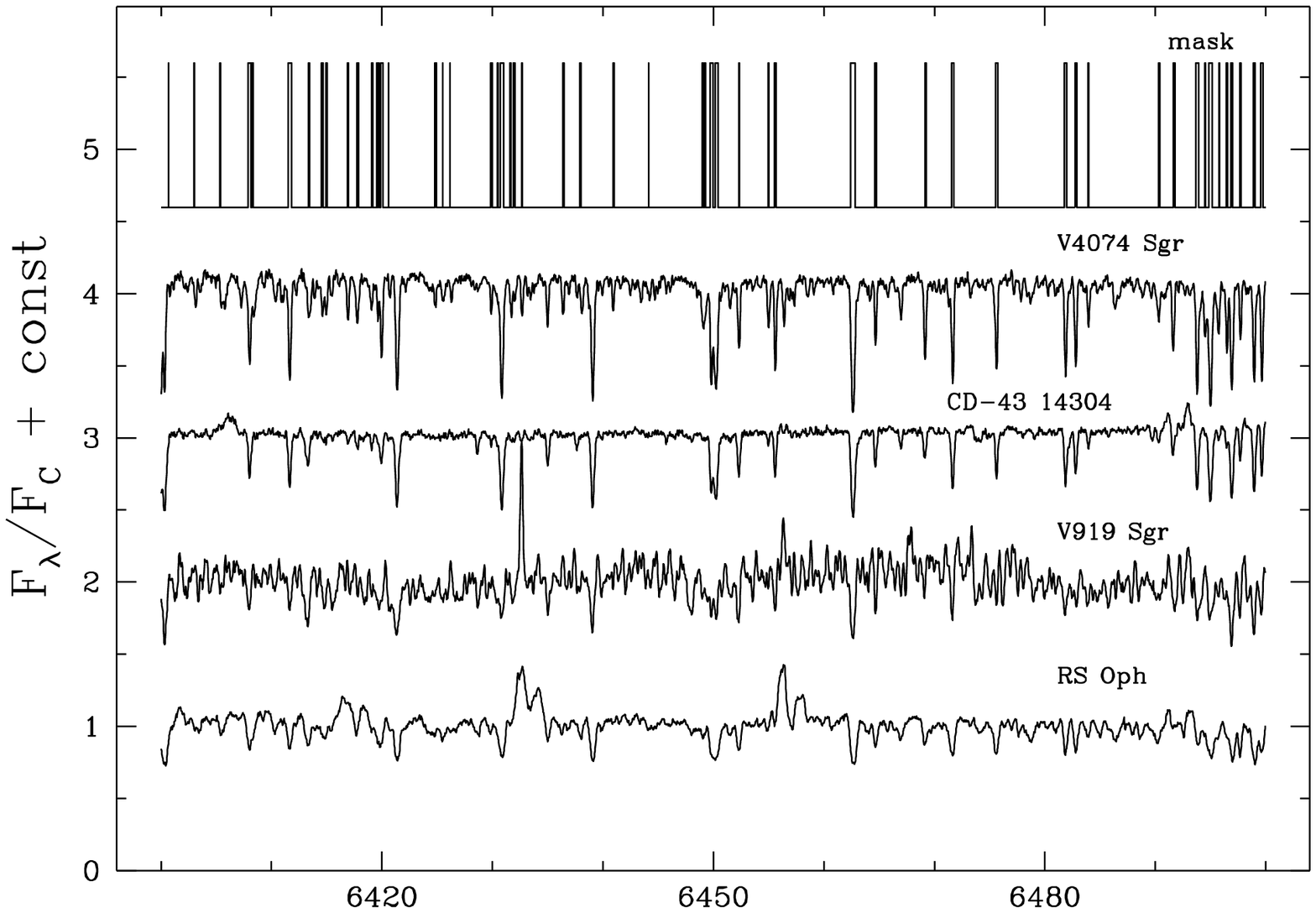}   
  \includegraphics{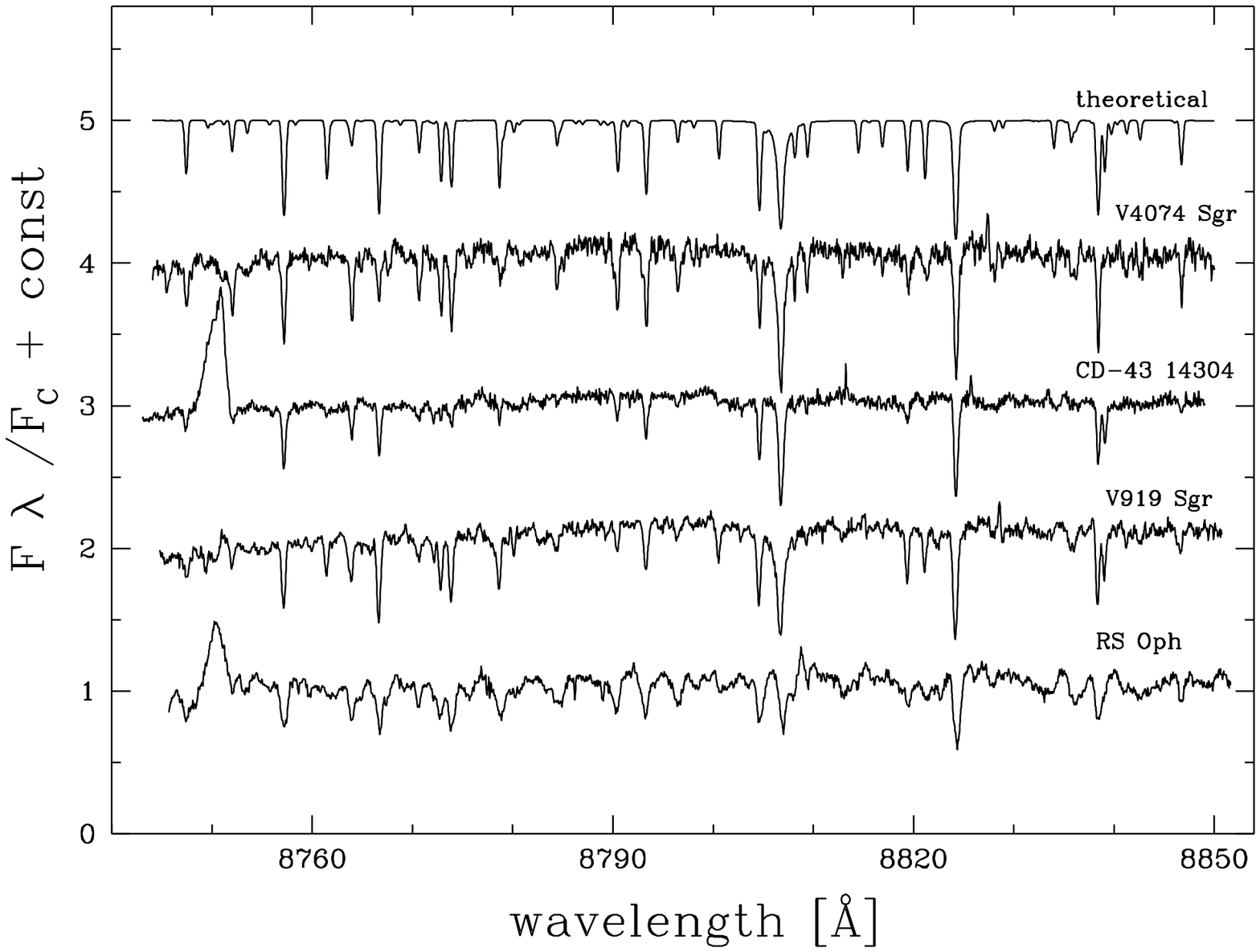}   
   \caption[]{A few examples of our spectra, together
   with the mask in the interval  $\lambda \lambda 6400 - 6500$ \AA (upper panel),
  and with a synthetic spectrum in the interval
  $\lambda \lambda 8740 - 8850$ \AA.
  The spectra are plotted in increasing order of rotation:
  V4074~Sgr (M4III, \vsi=3.5 \kms),
  CD-43$^{\circ}$14304 (K5III, \vsi=7.2 \kms),
  V919~Sgr (M2III, \vsi=9.1 \kms),
  RS~Oph (M0III, \vsi=11.7 \kms).
  The synthetic spectrum corresponds to M0III and \vsi=5 \kms. 
  }		    
\label{figEx}     
\end{figure*}

Here we give a few examples of our spectra (Fig.~\ref{figEx}) and 
graphic representation of the CCF for all SSs observed in this paper
(Fig.\ref{CCF}).

 \begin{figure*}   
 \mbox{}   
 \vspace{20.0cm}   
  \includegraphics{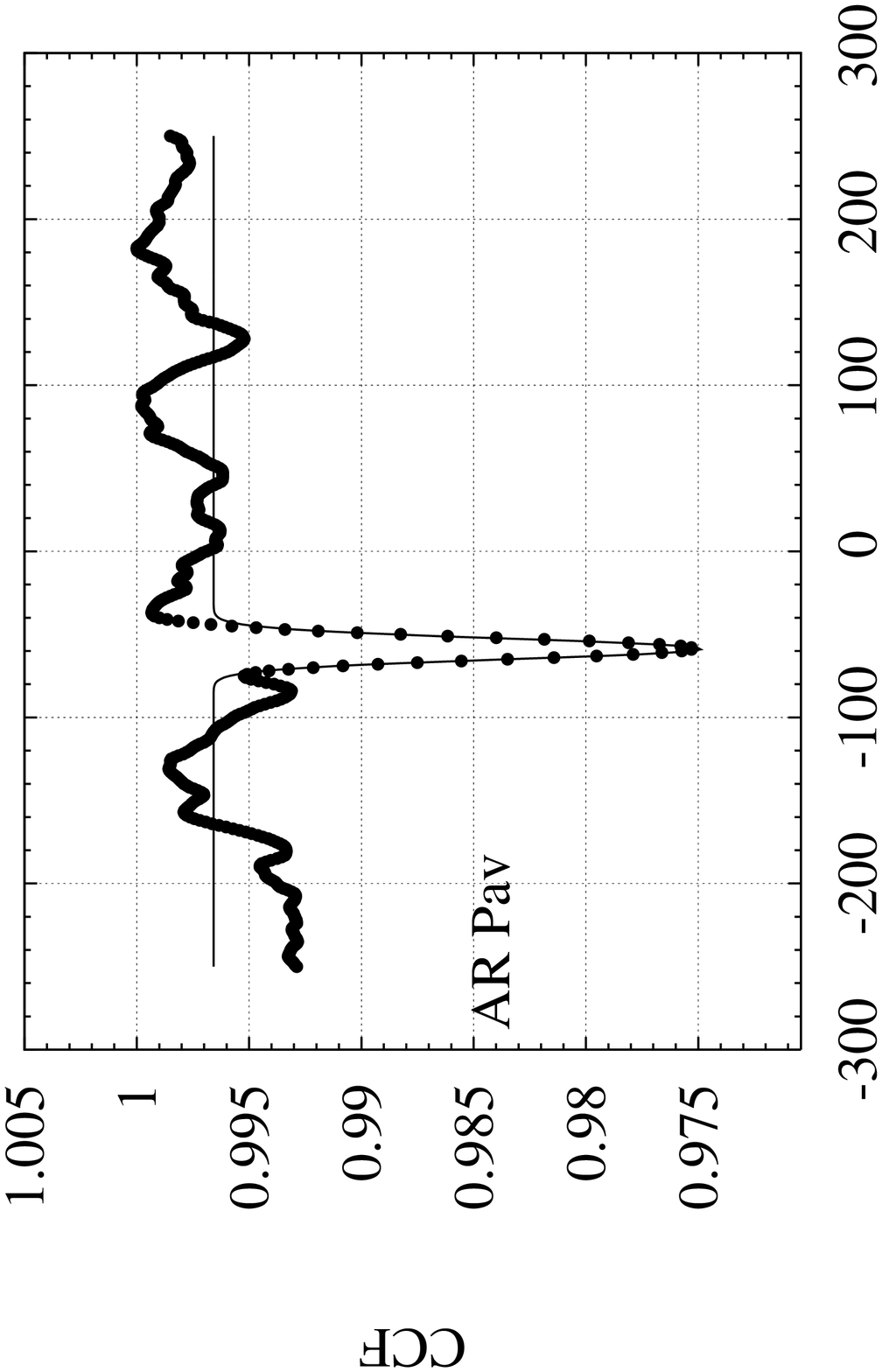}    
  \includegraphics{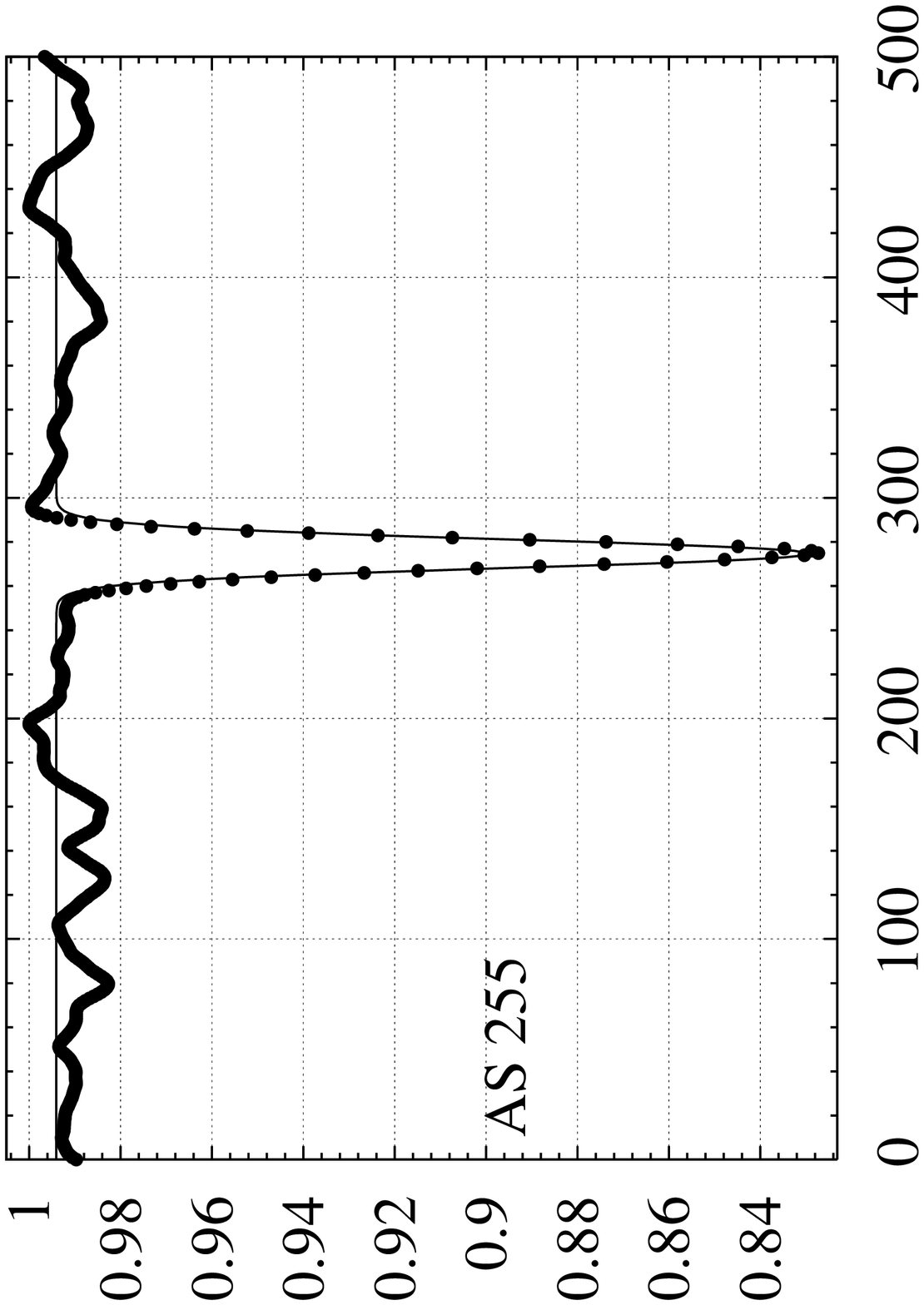}    
  \includegraphics{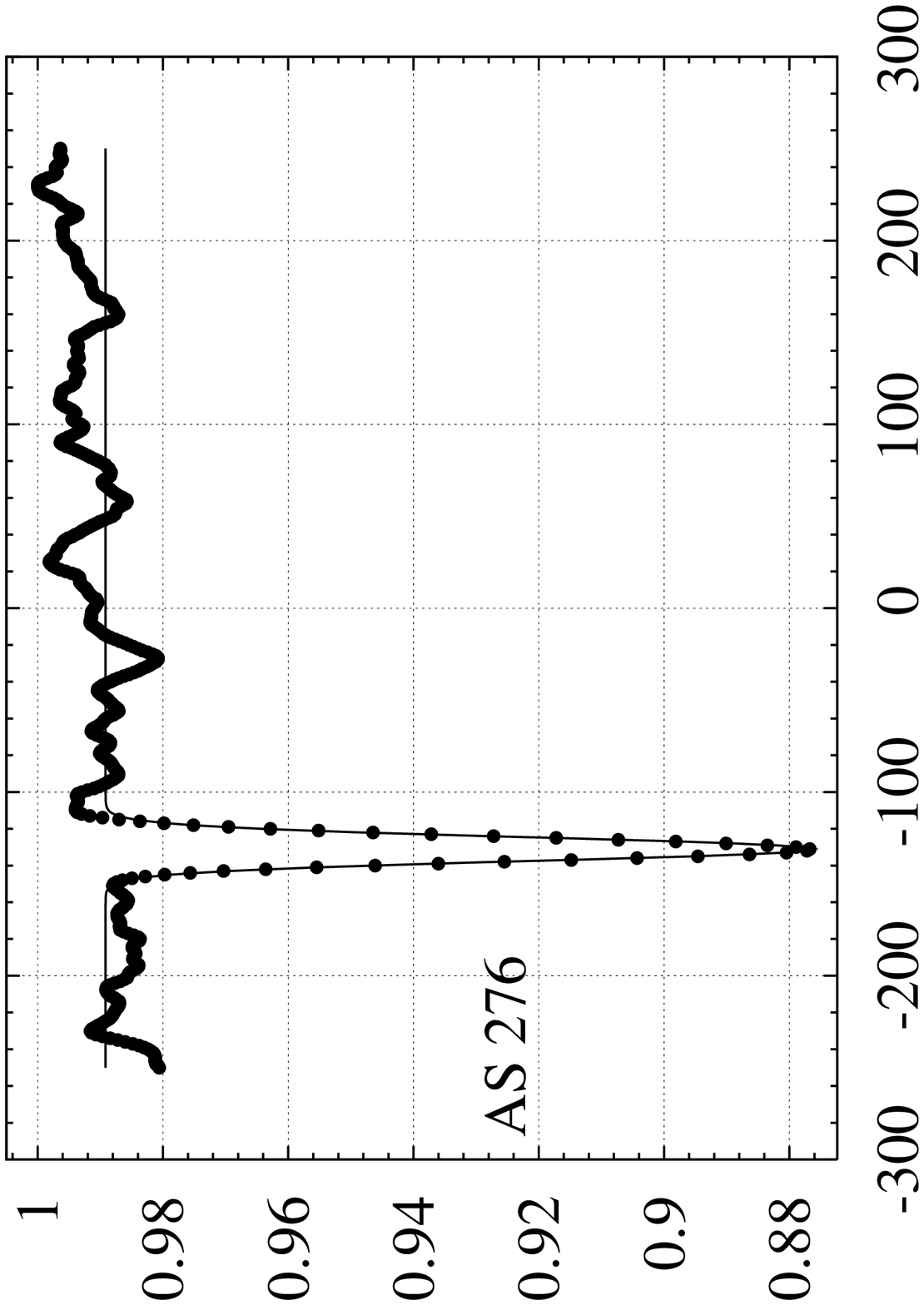}   
  \includegraphics{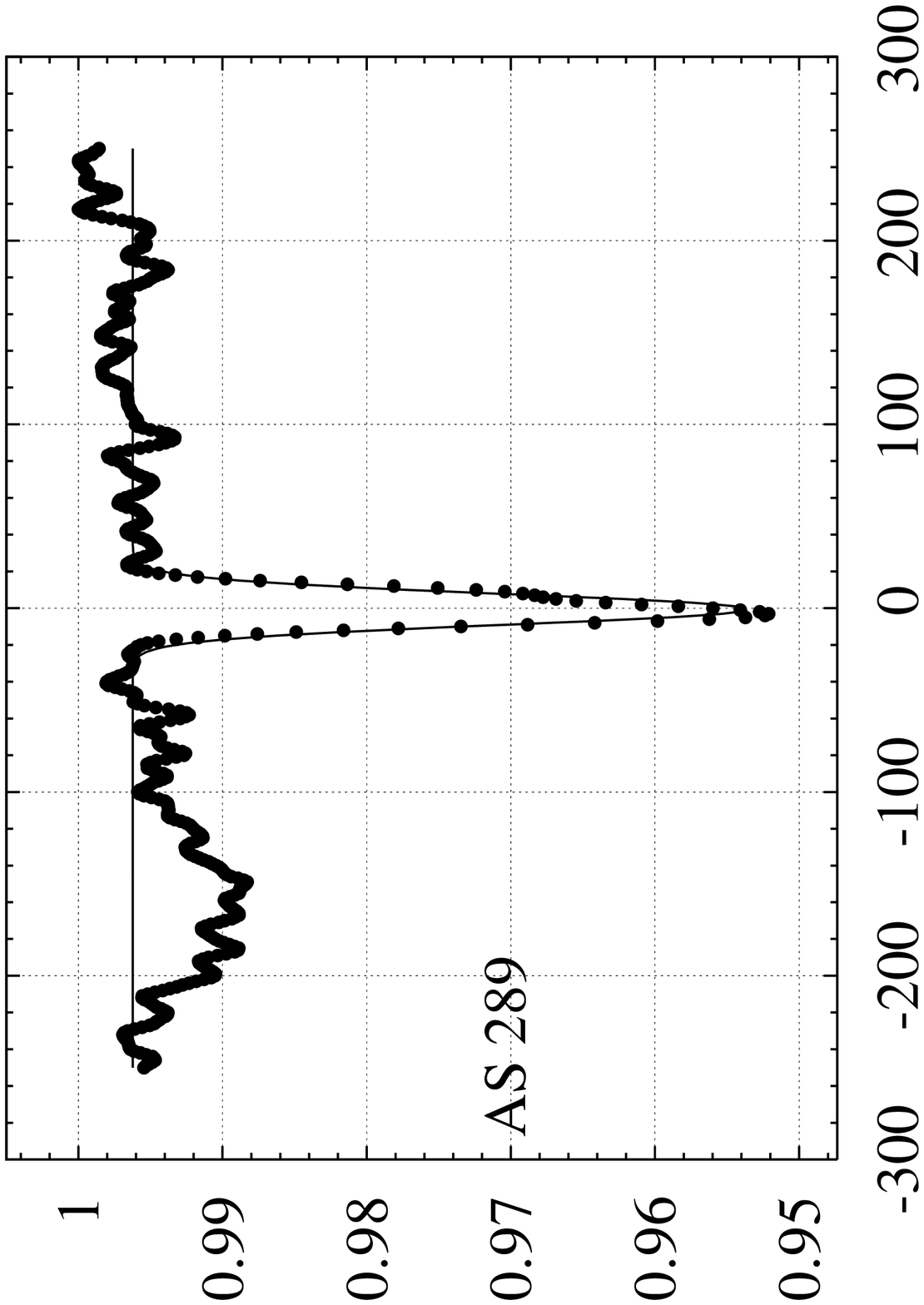}    
  \includegraphics{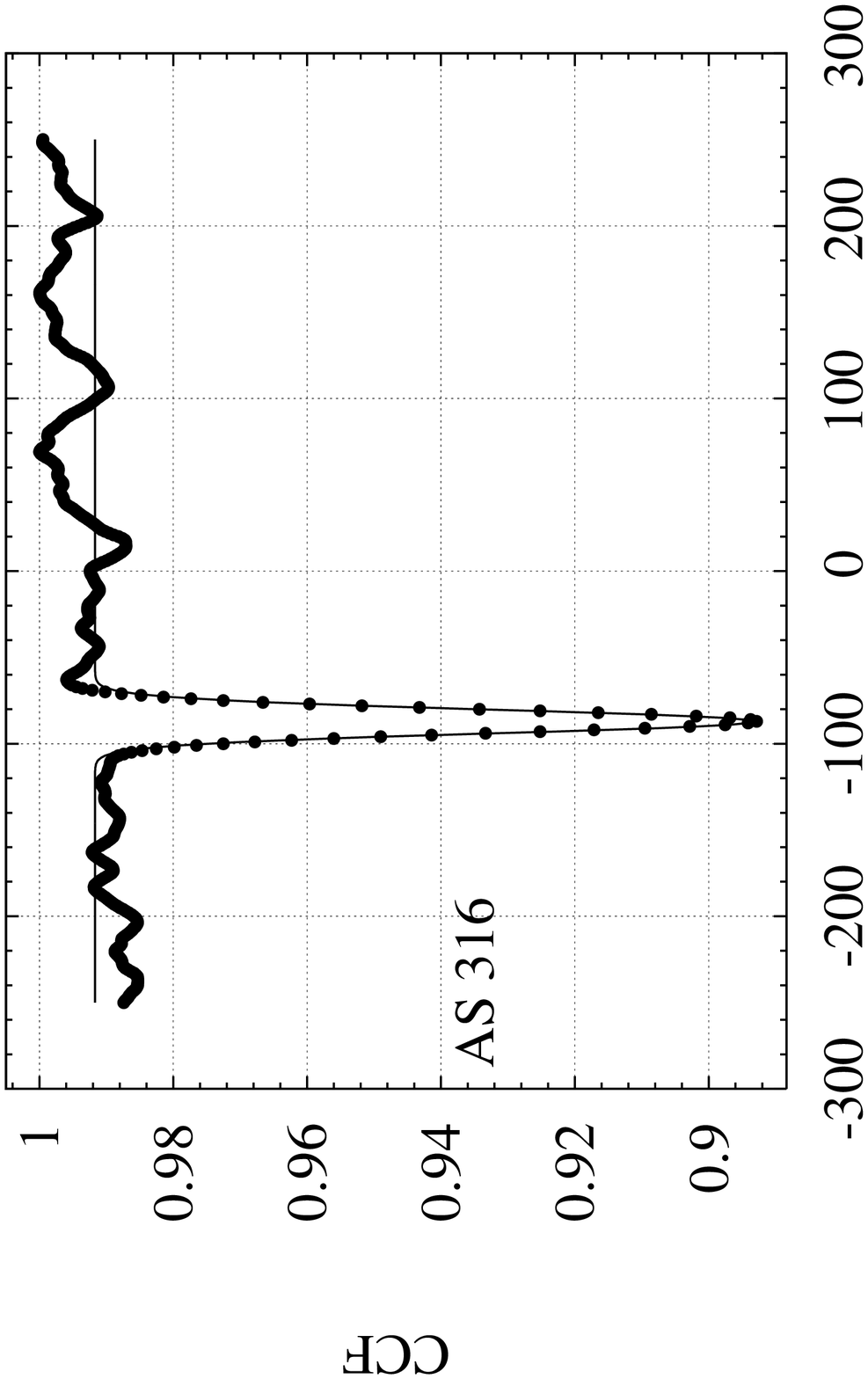} 
  \includegraphics{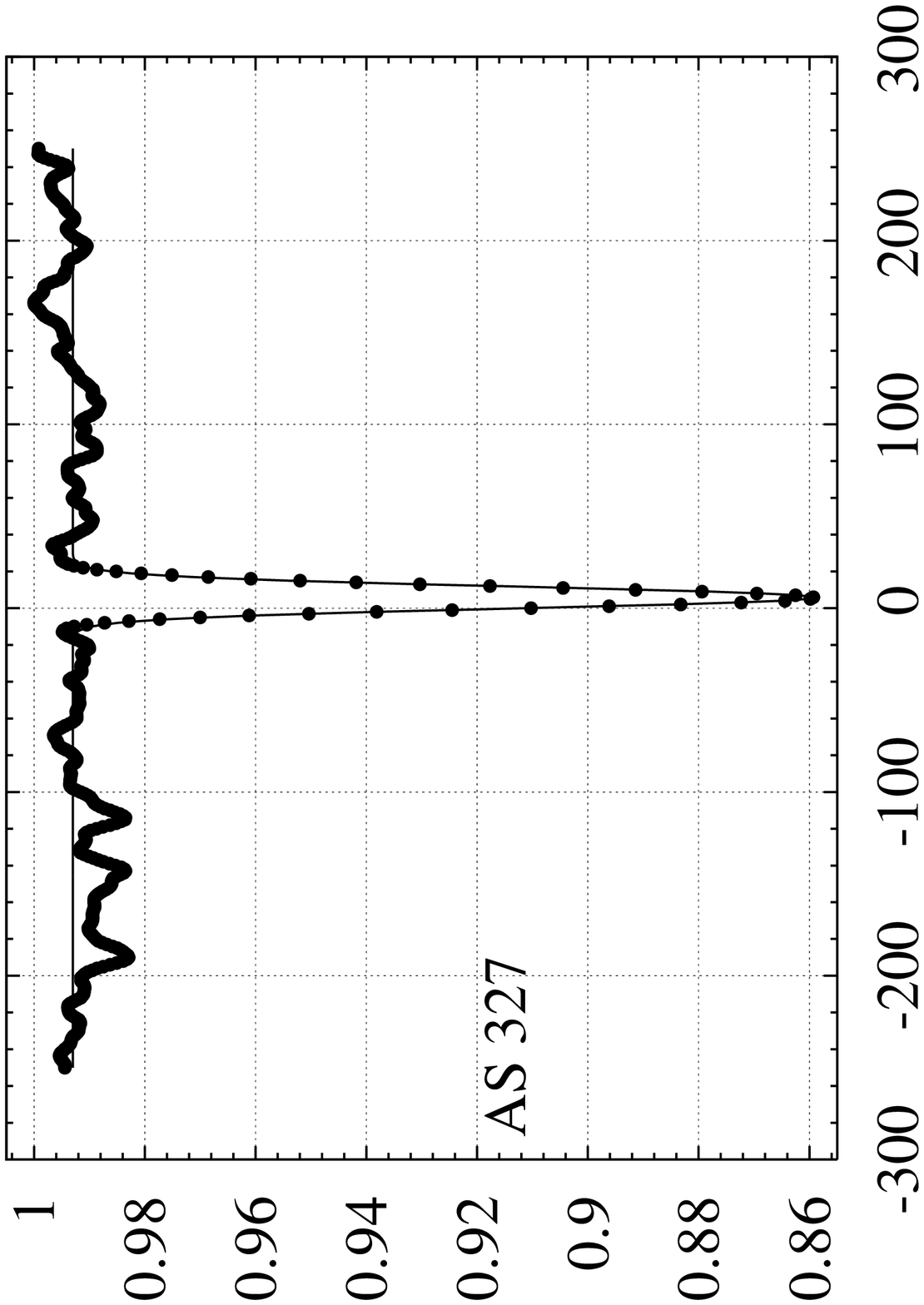} 
  \includegraphics{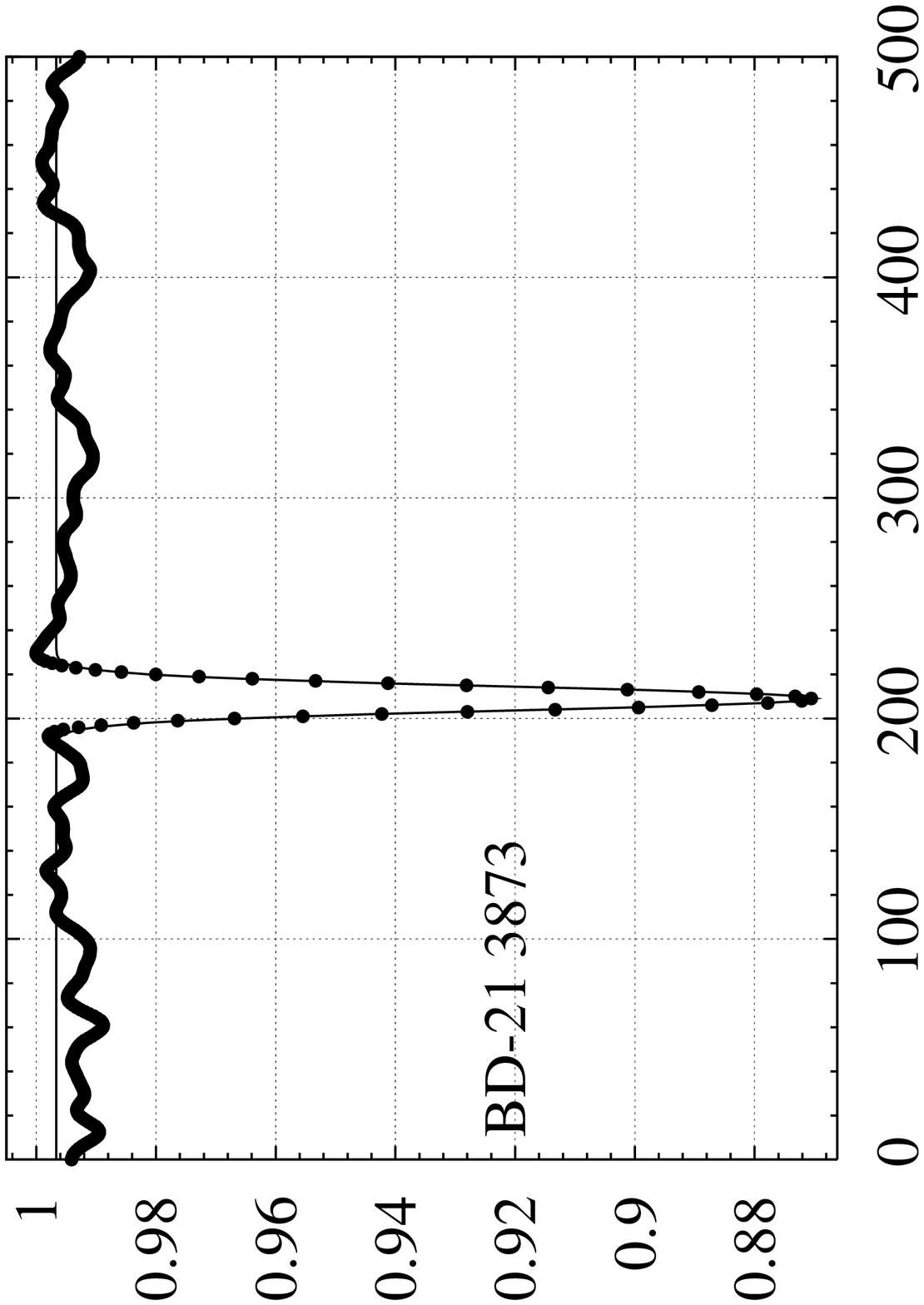}   
  \includegraphics{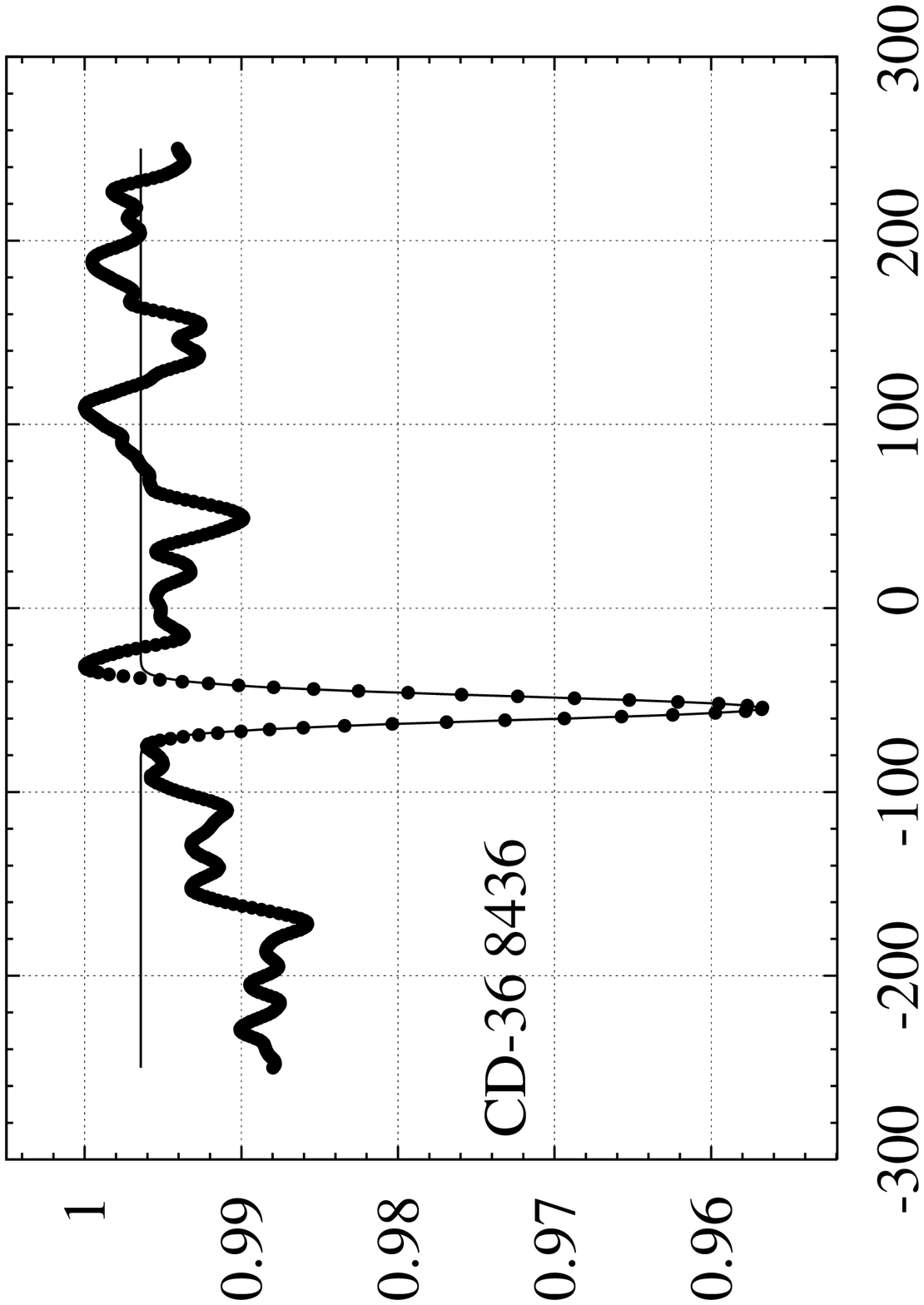}  
  \includegraphics{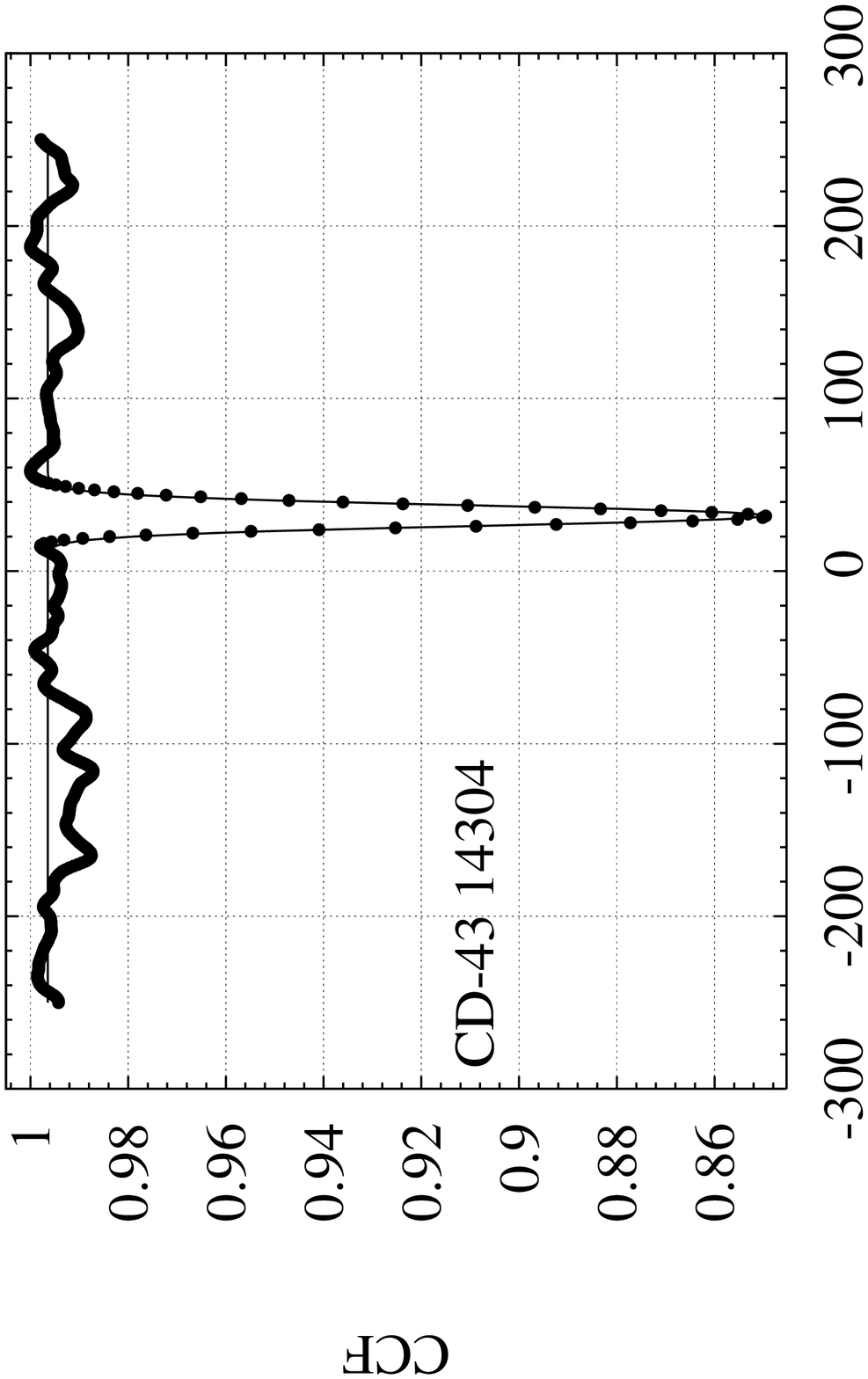}  
  \includegraphics{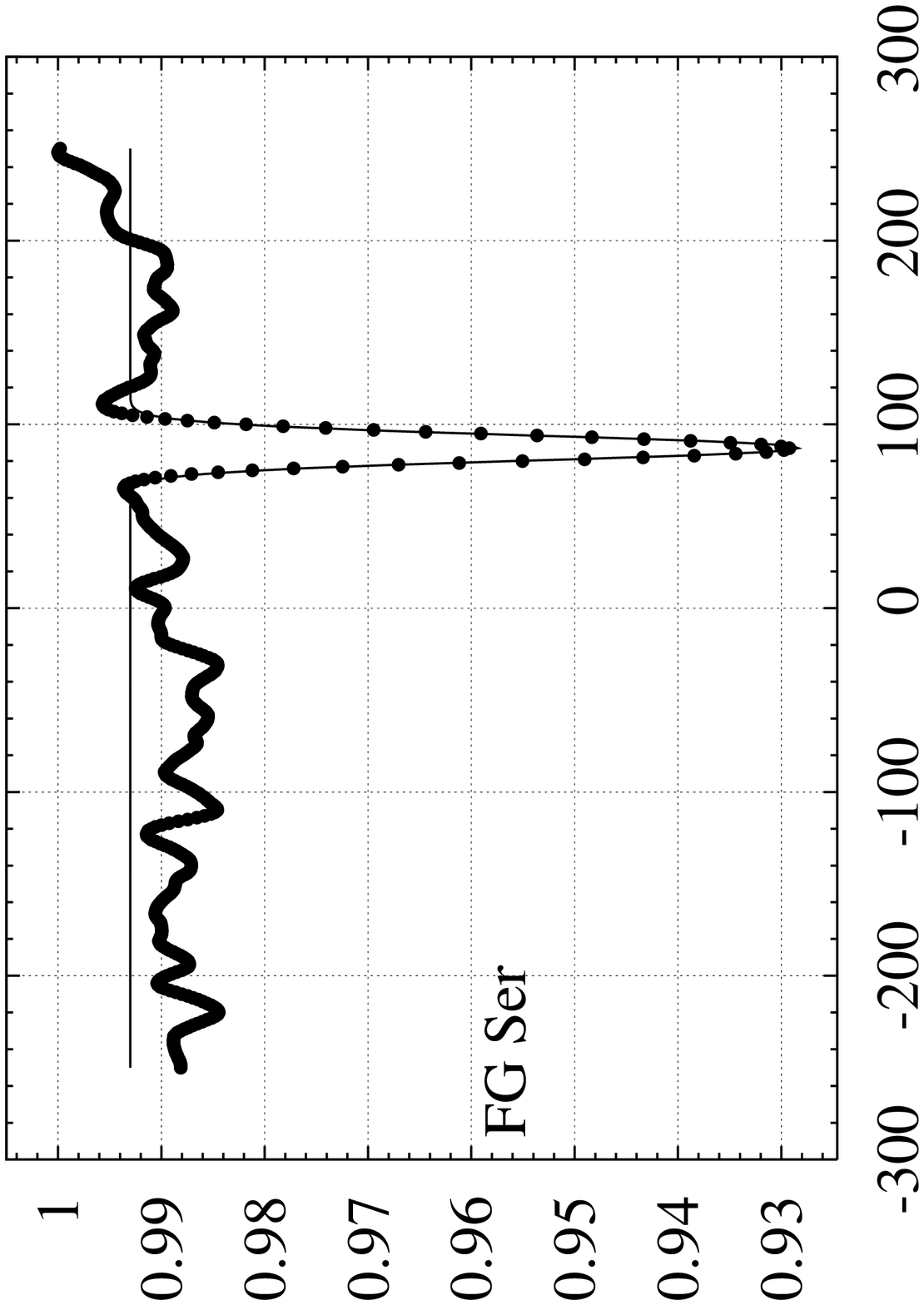}  
  \includegraphics{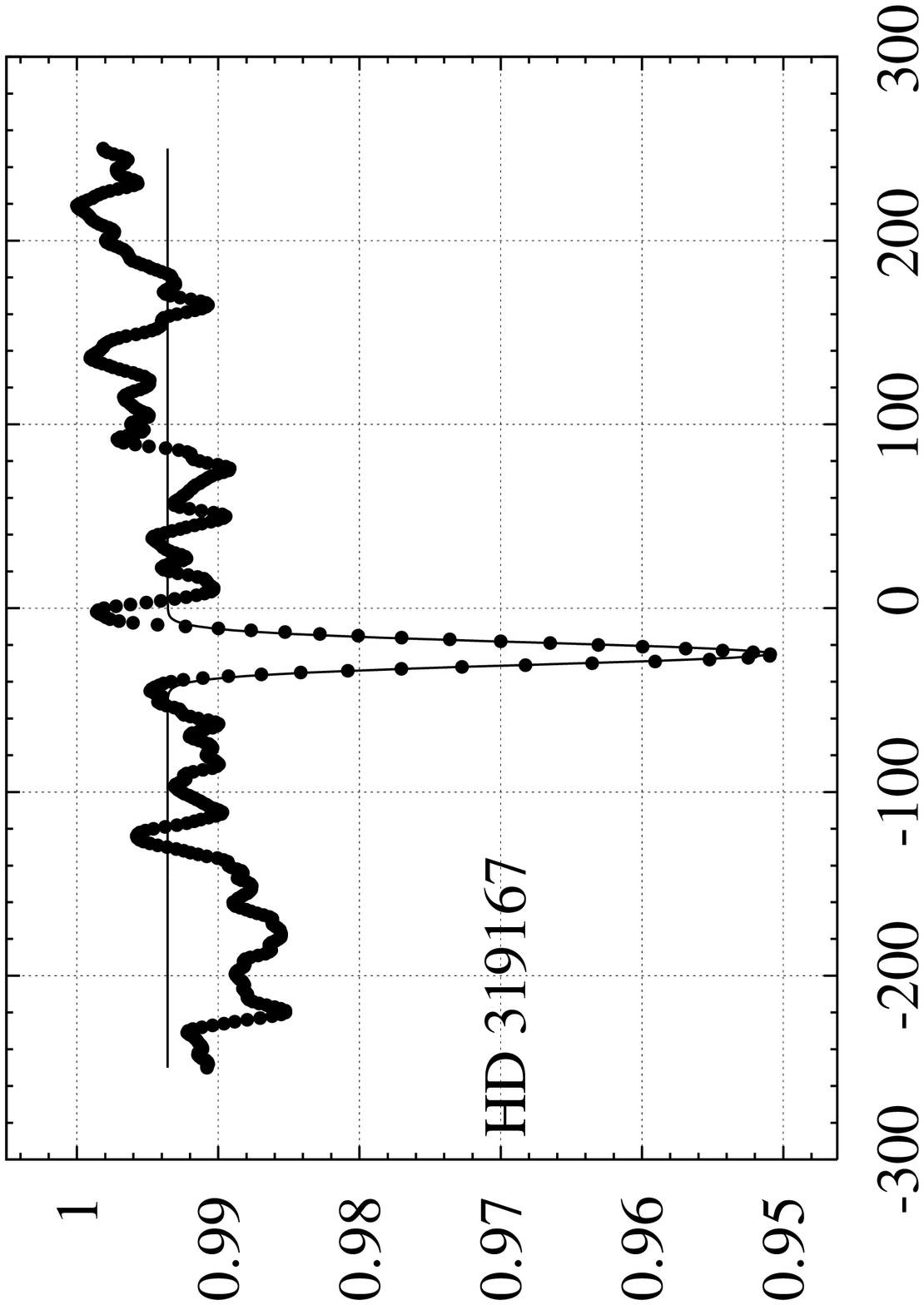}   
  \includegraphics{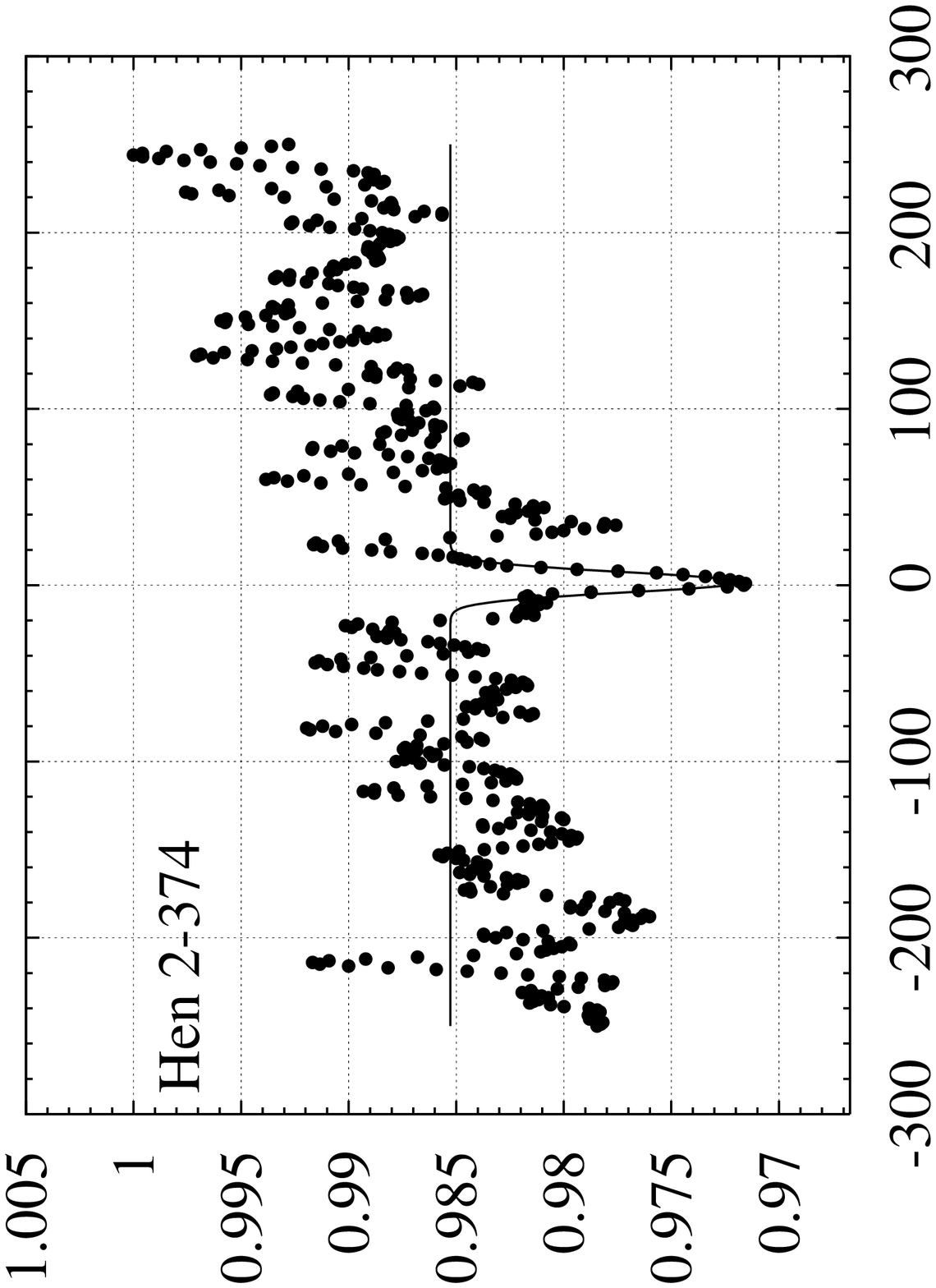} 
  \includegraphics{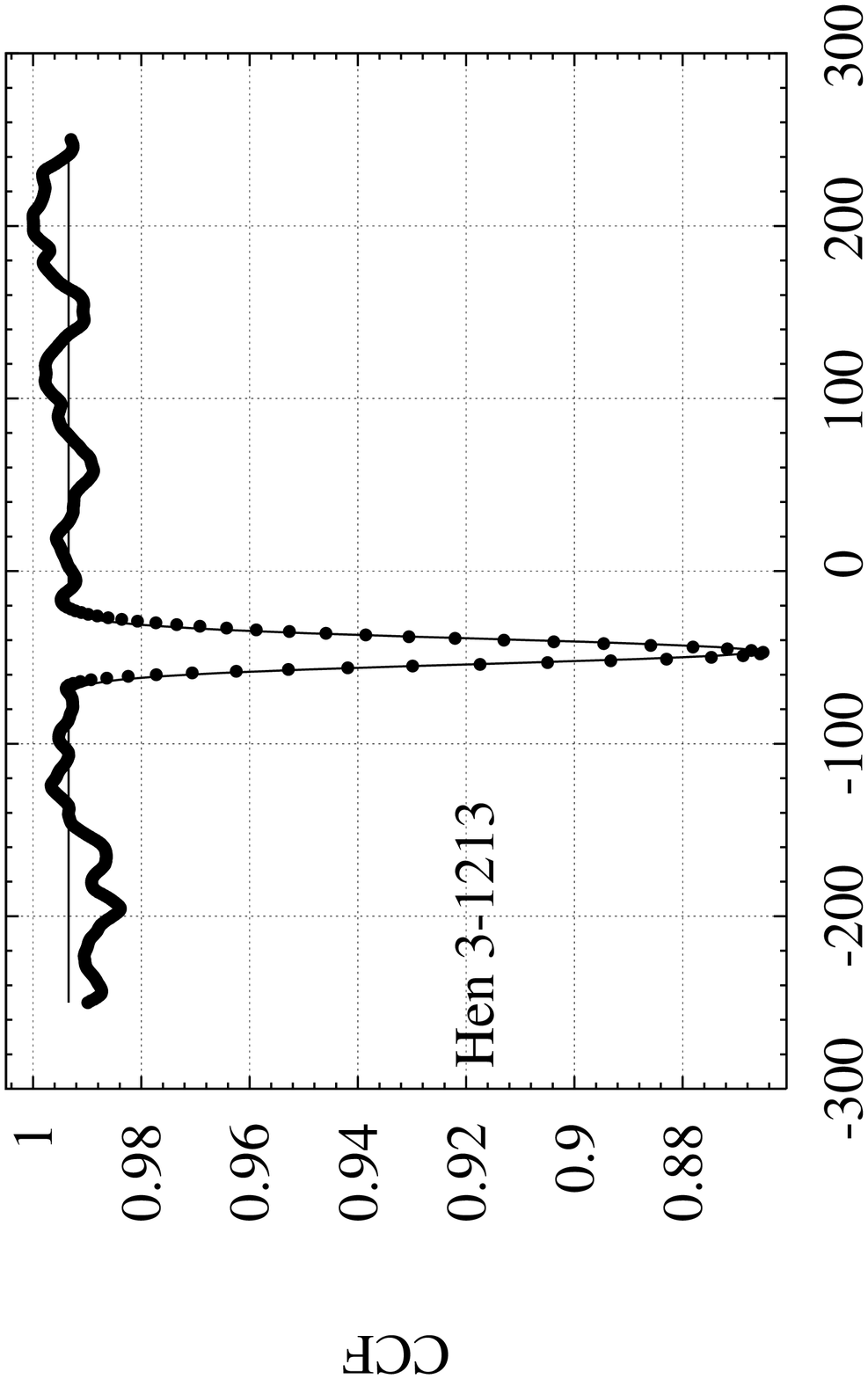}  
  \includegraphics{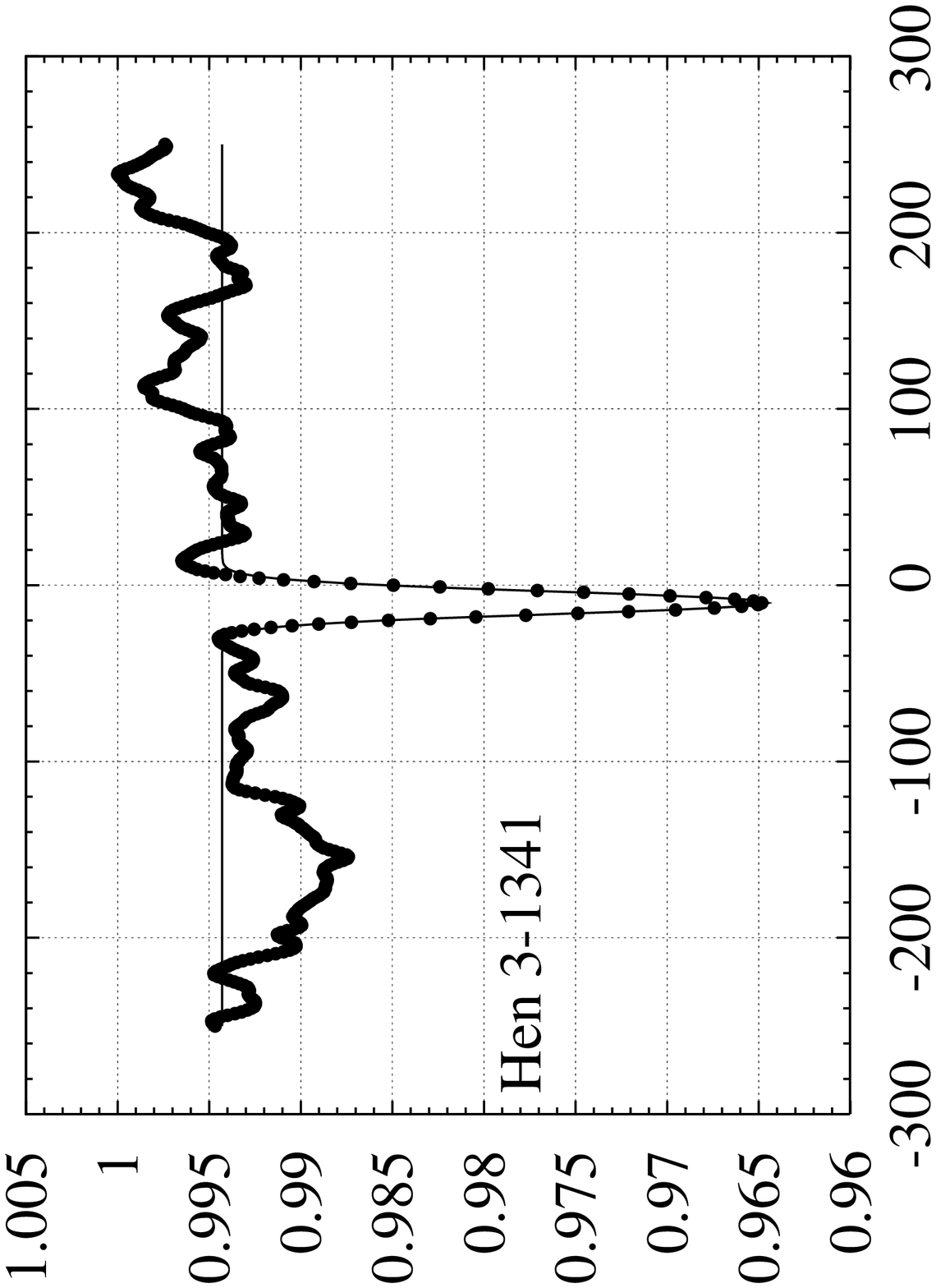}  
  \includegraphics{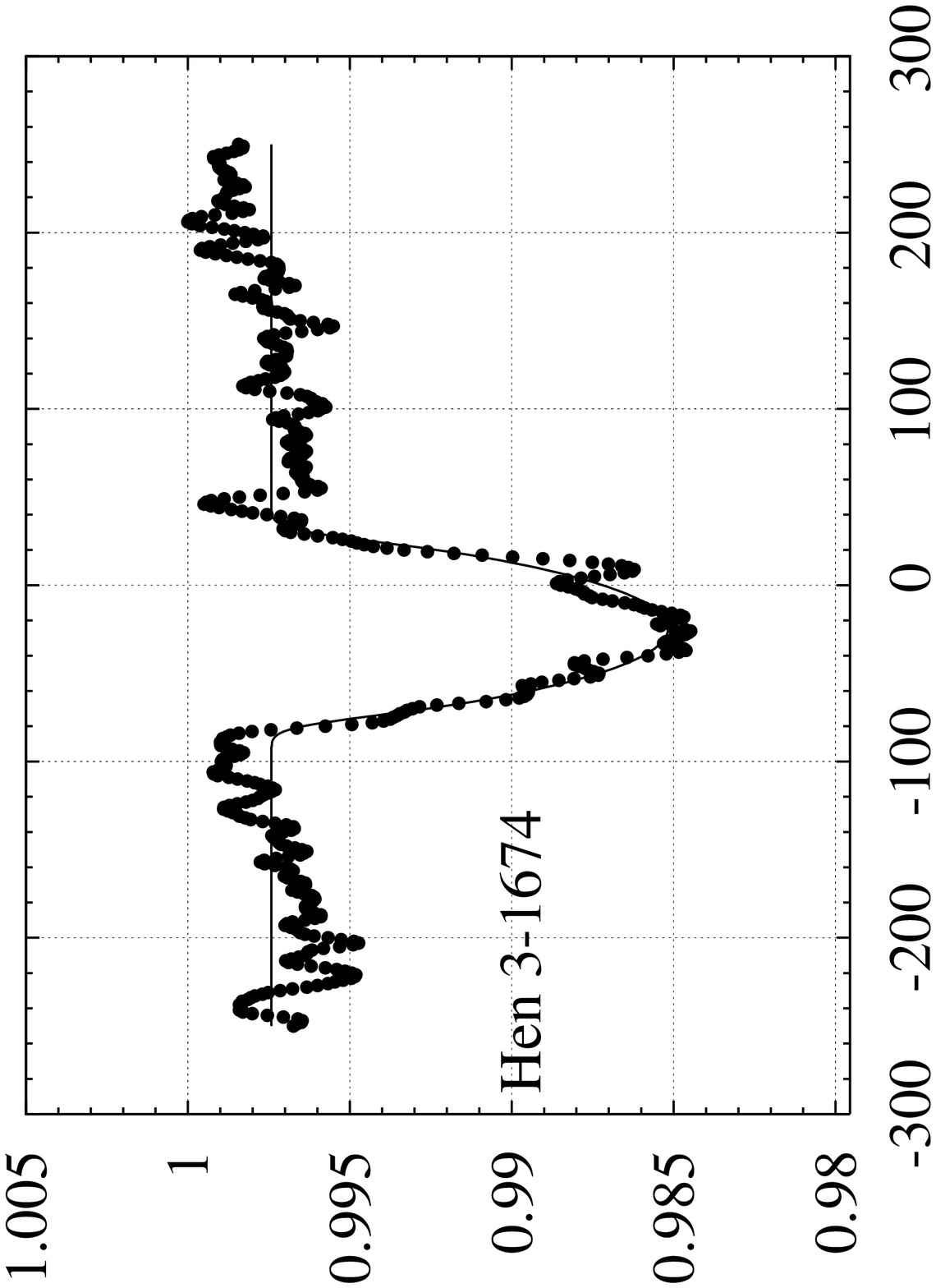}   
  \includegraphics{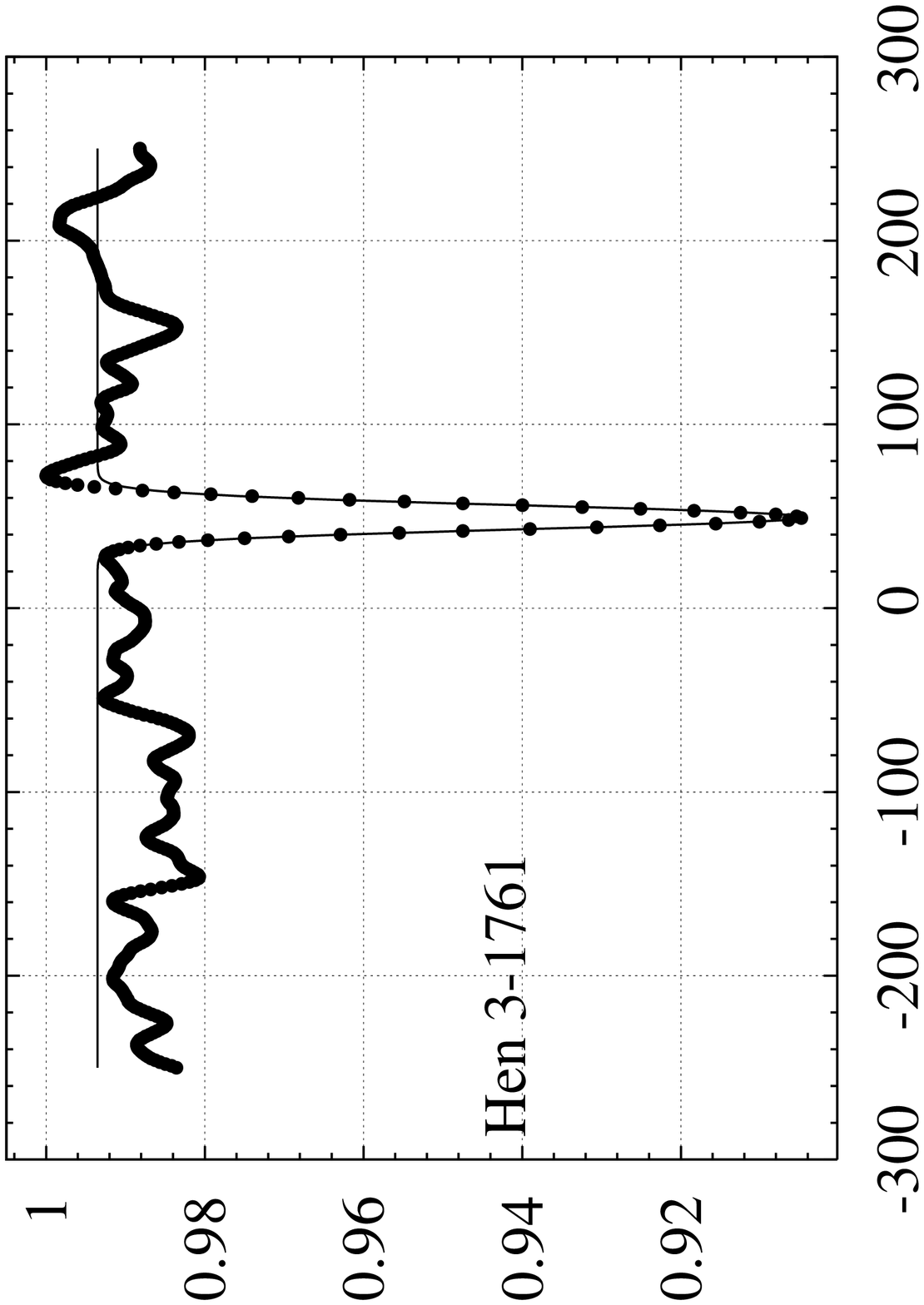} 
  \includegraphics{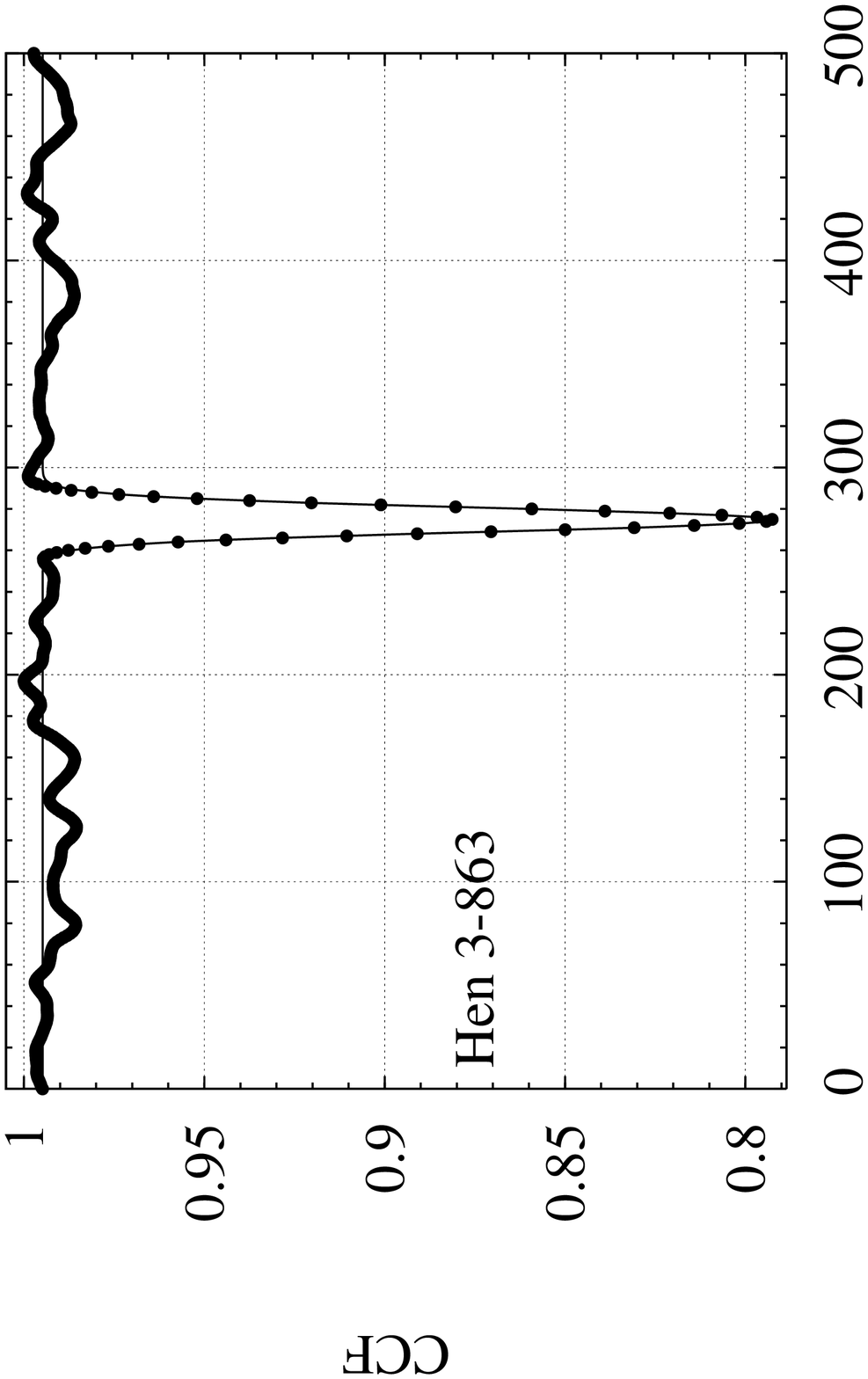}	
  \includegraphics{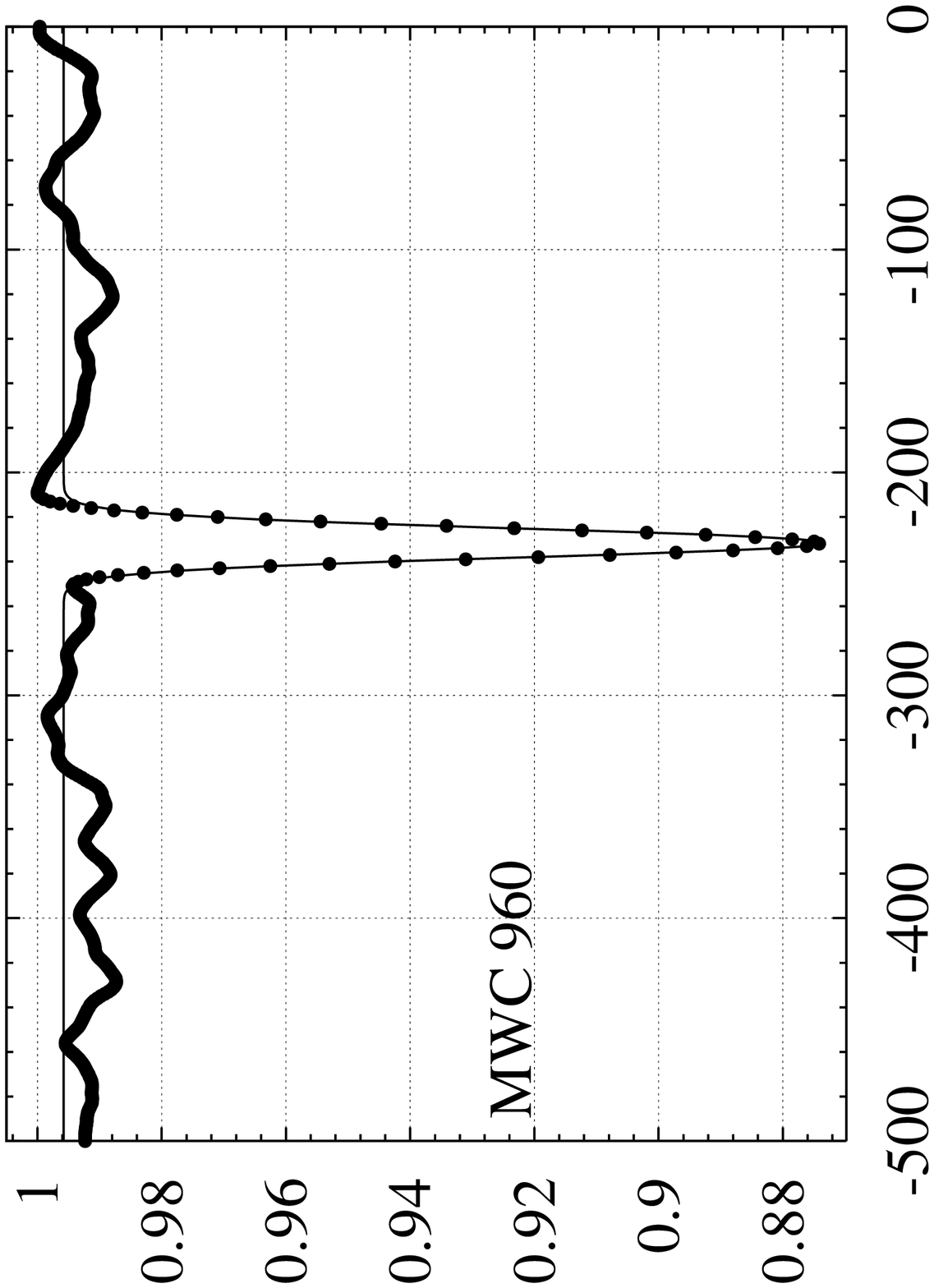}	
  \includegraphics{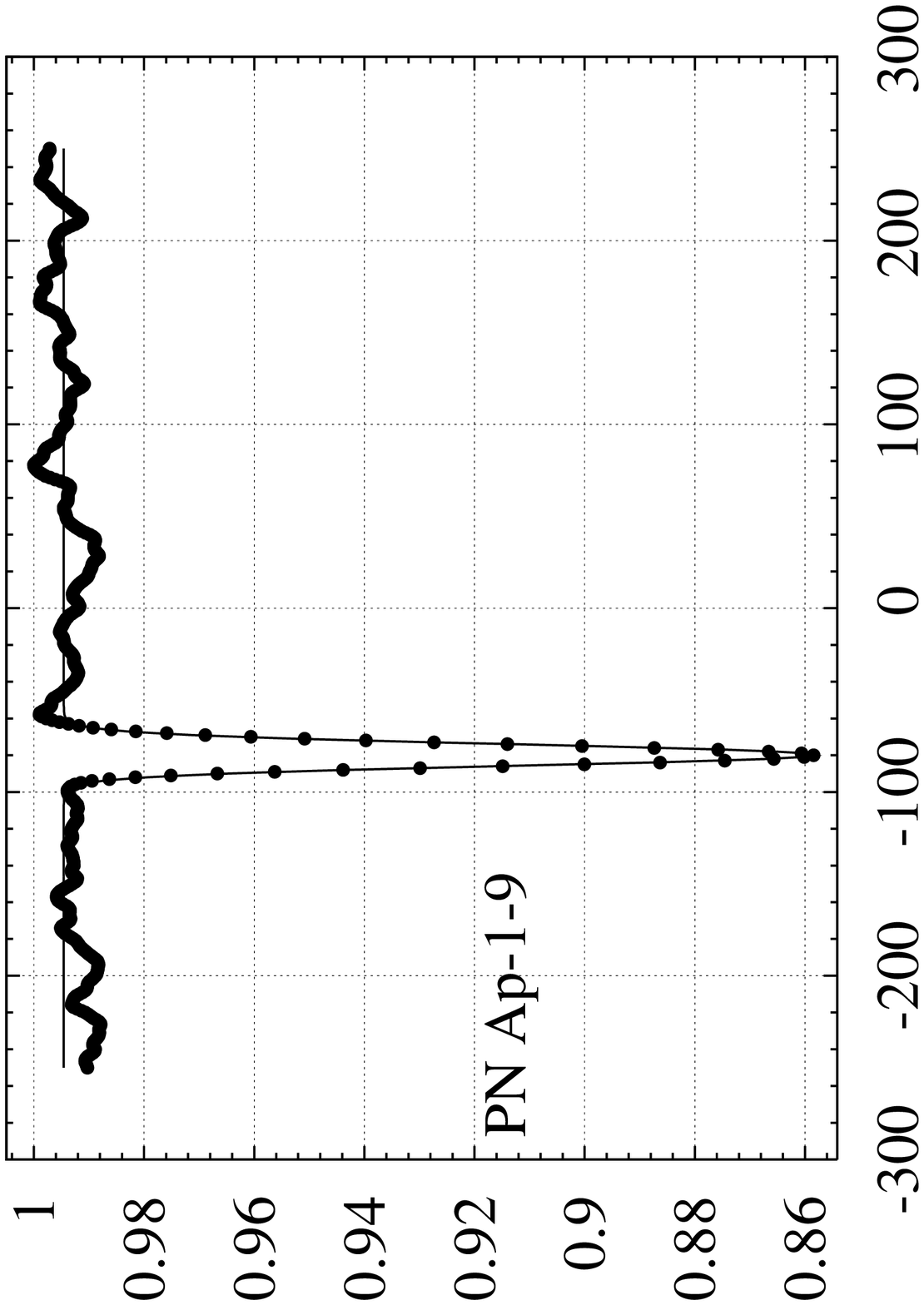}   
  \includegraphics{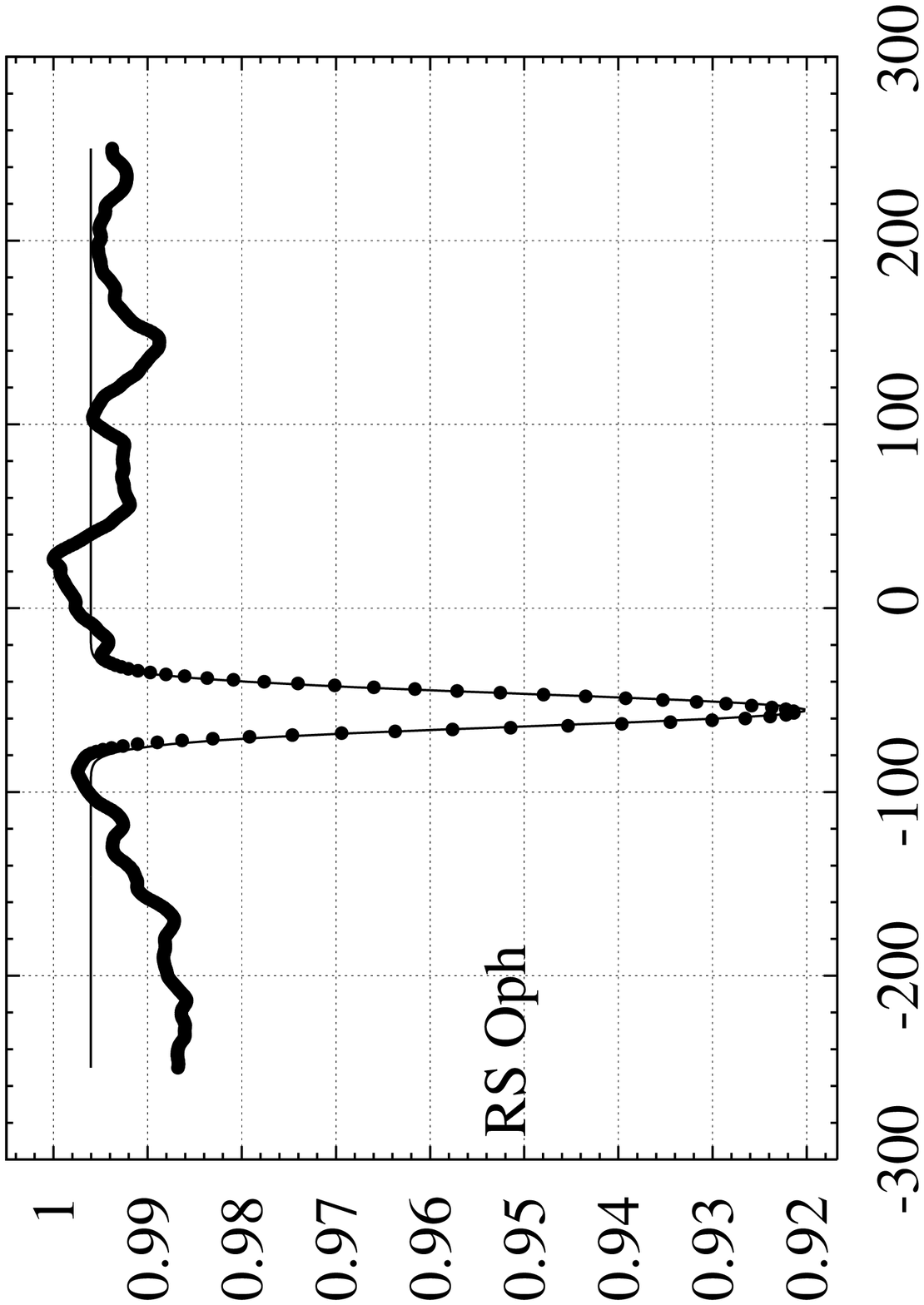}	
  \includegraphics{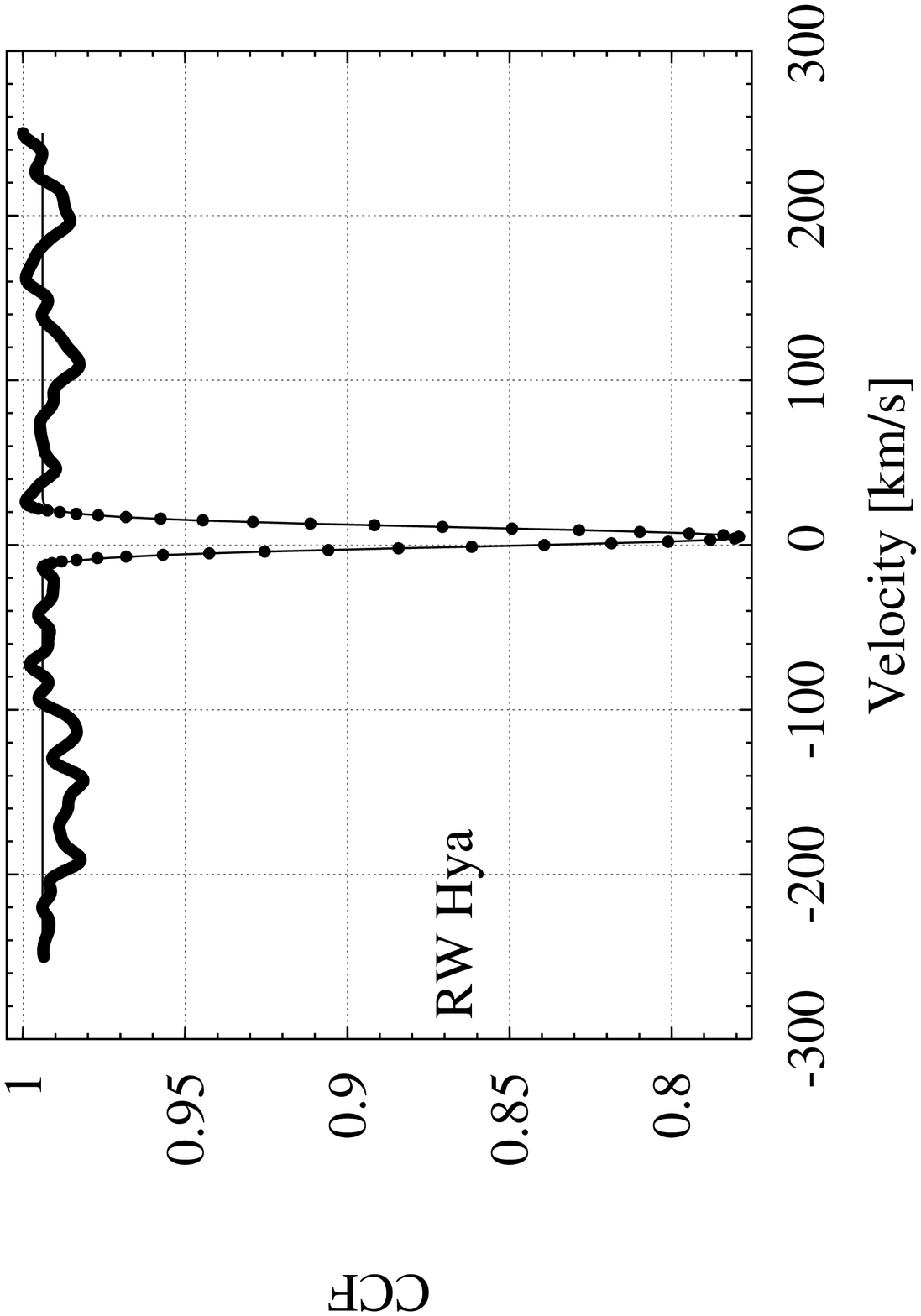}  
  \includegraphics{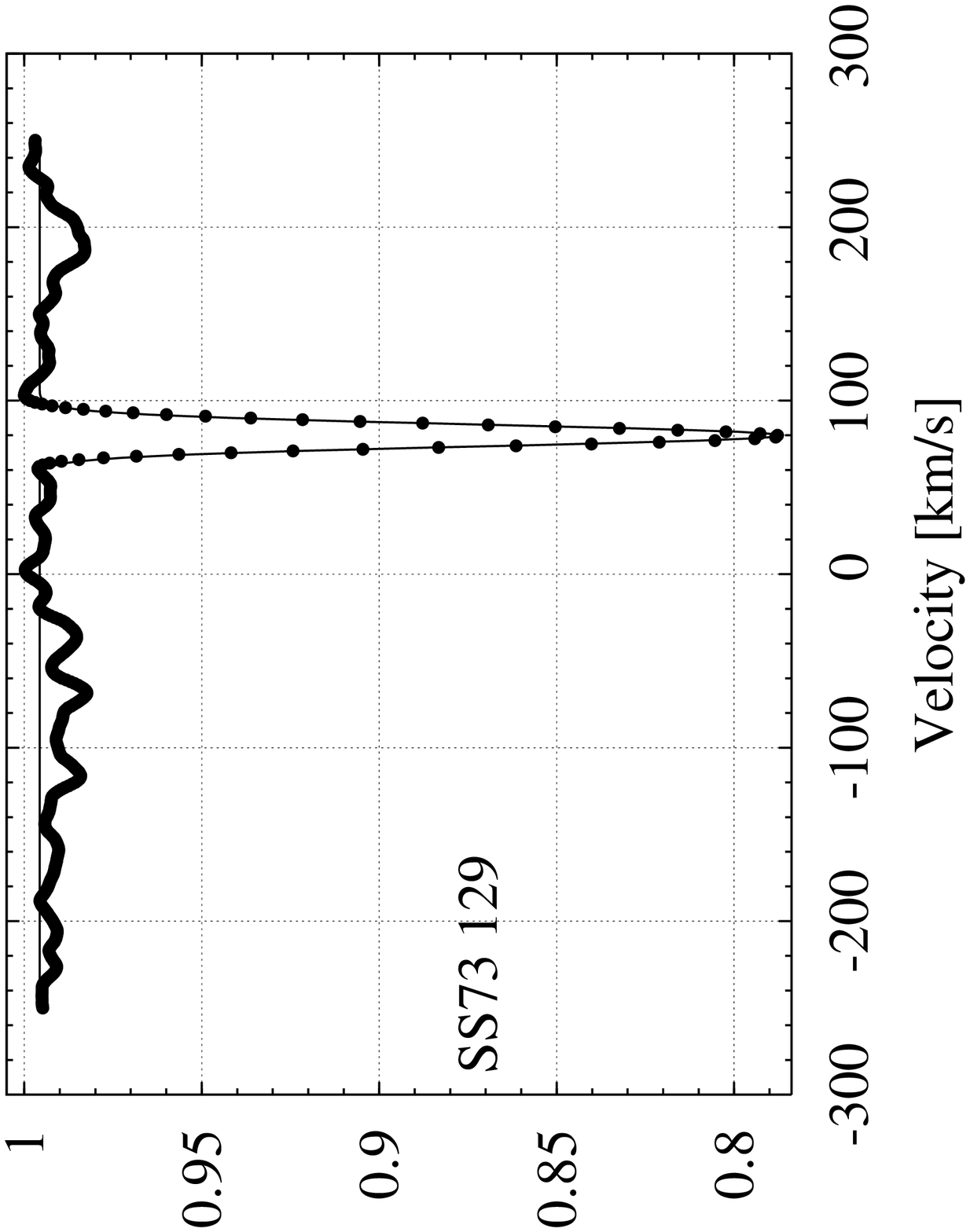}  
  \includegraphics{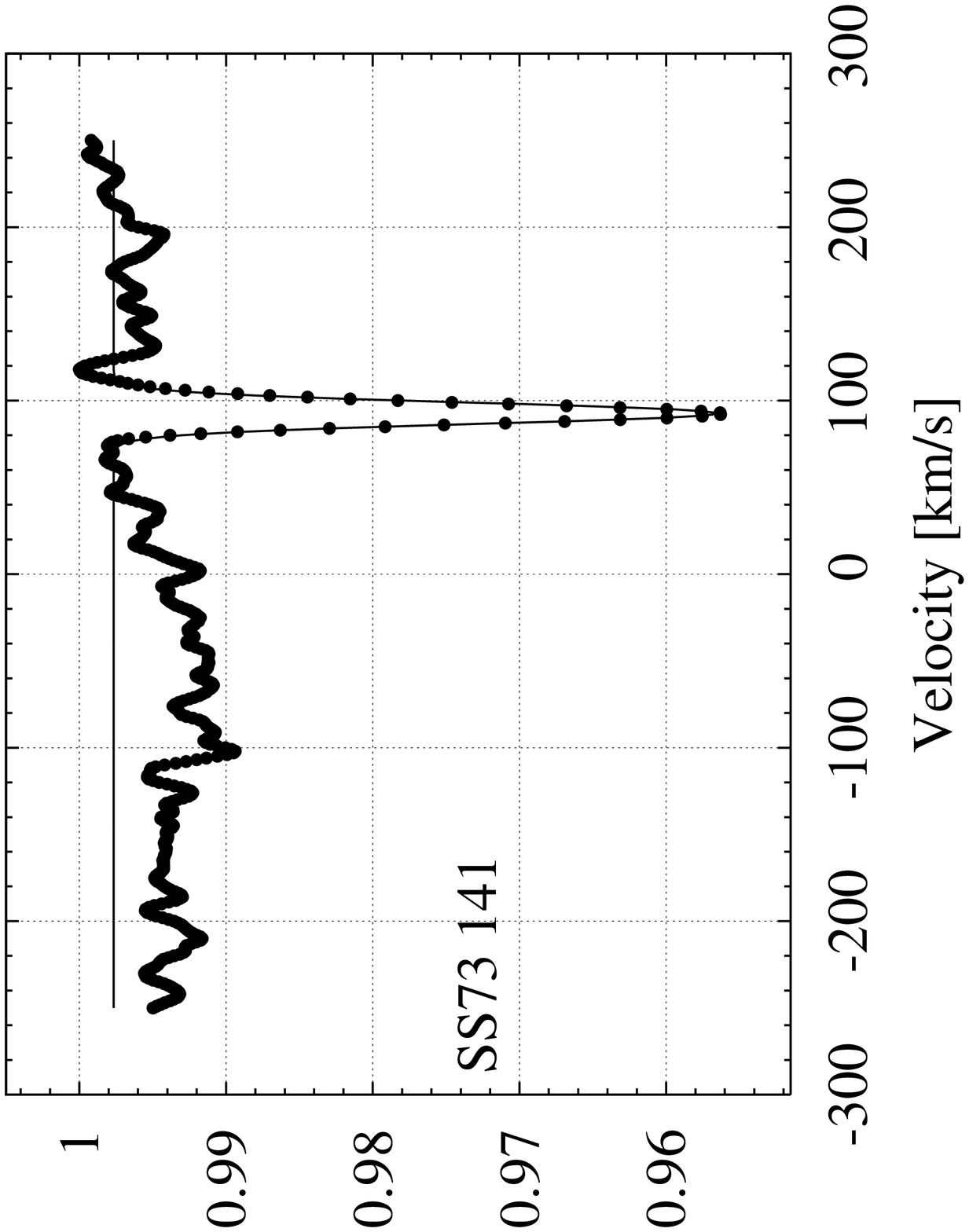}   
  \includegraphics{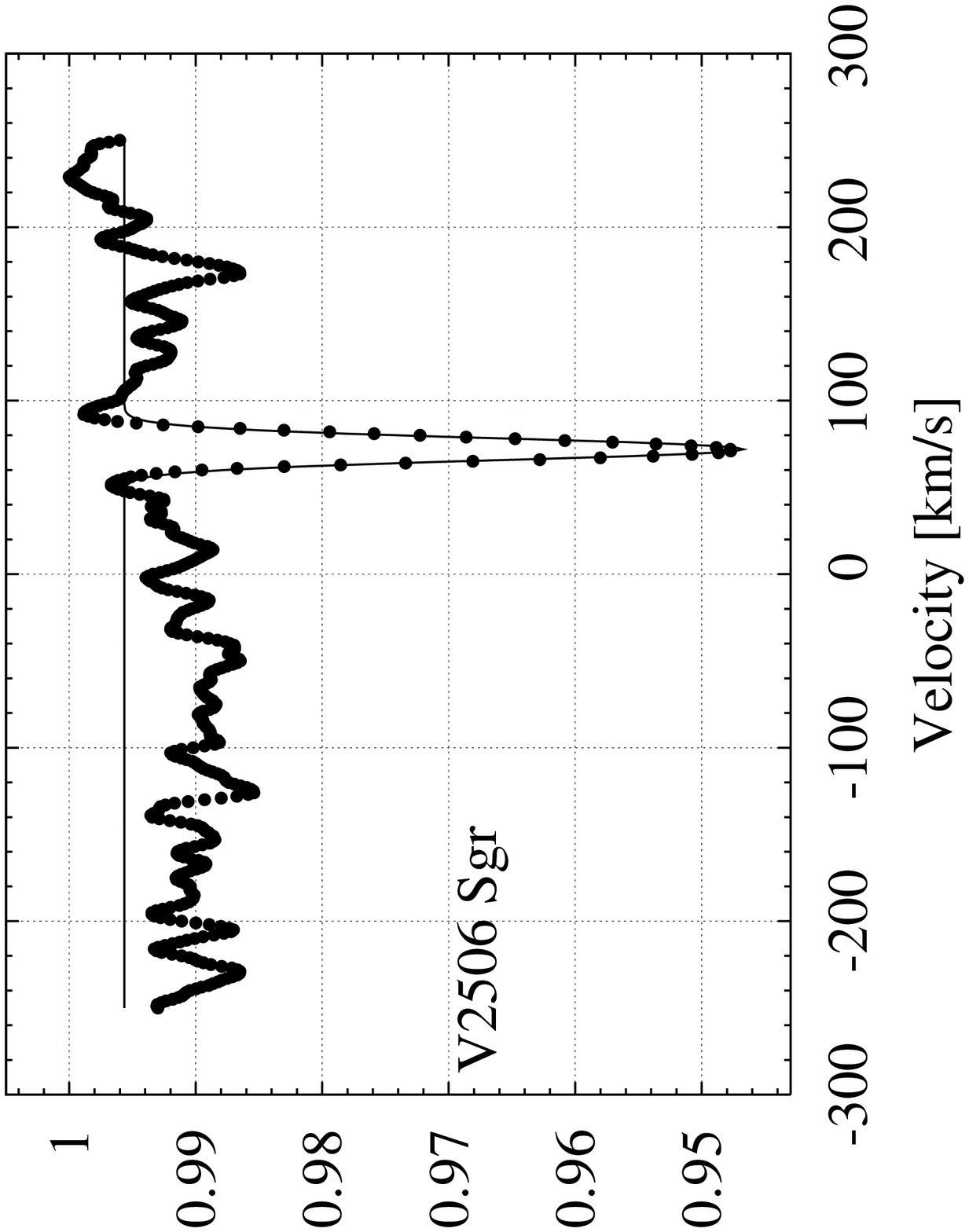} 
\caption[]{ CCF using K0 numerical mask. In each panel are plotted the relative intensity 
            of CCF (heavy line) and the fit versus 
            radial velocity for the SSs observed in this paper. The measured
	    widths of the CCF are given in Table~\ref{tab_vsi}. }	     
\label{CCF}      
\end{figure*}	
\setcounter{figure}{4}
 \begin{figure*}   
 \mbox{}   
 \vspace{7.0cm}   
  \includegraphics{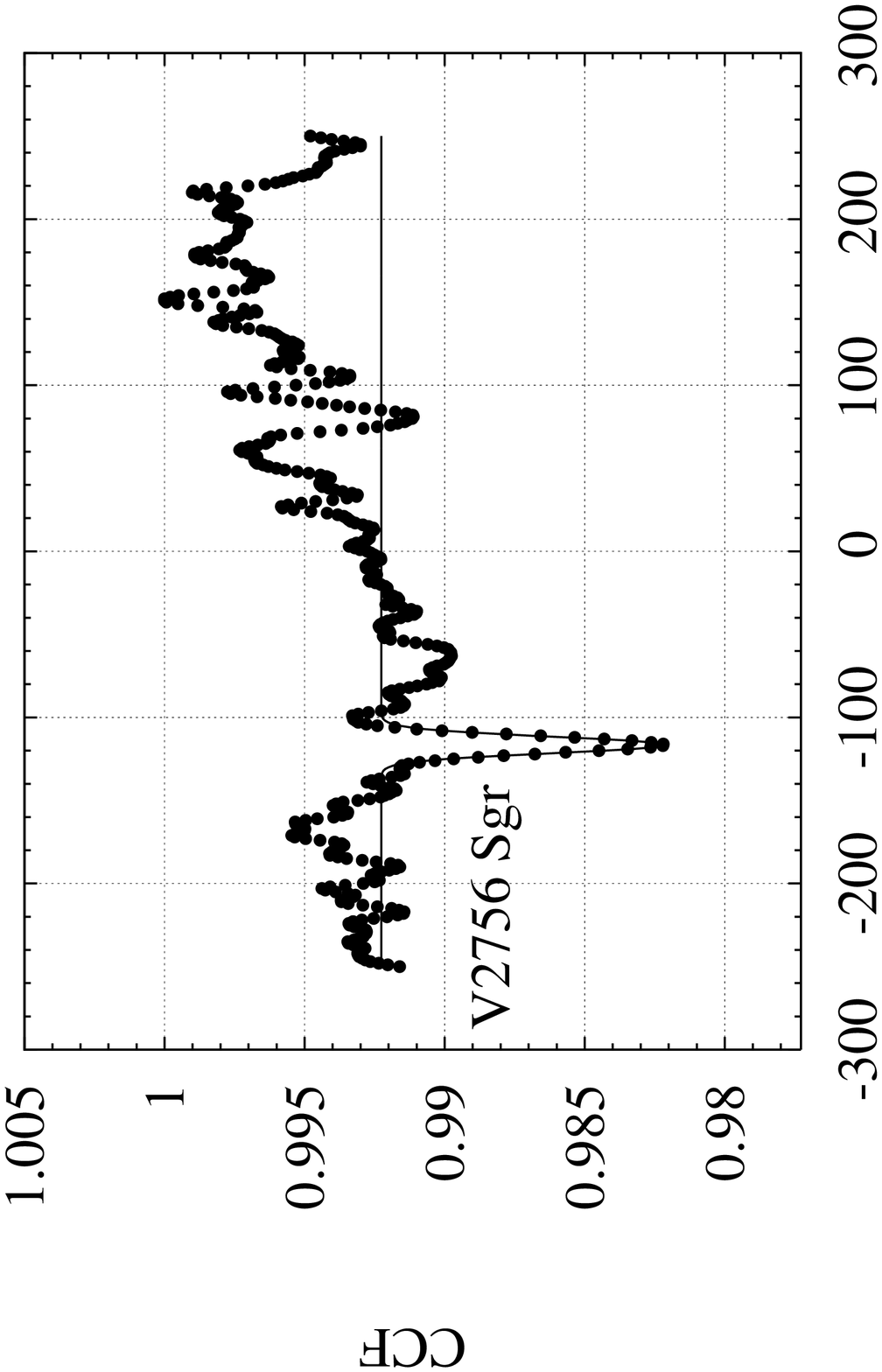}	
  \includegraphics{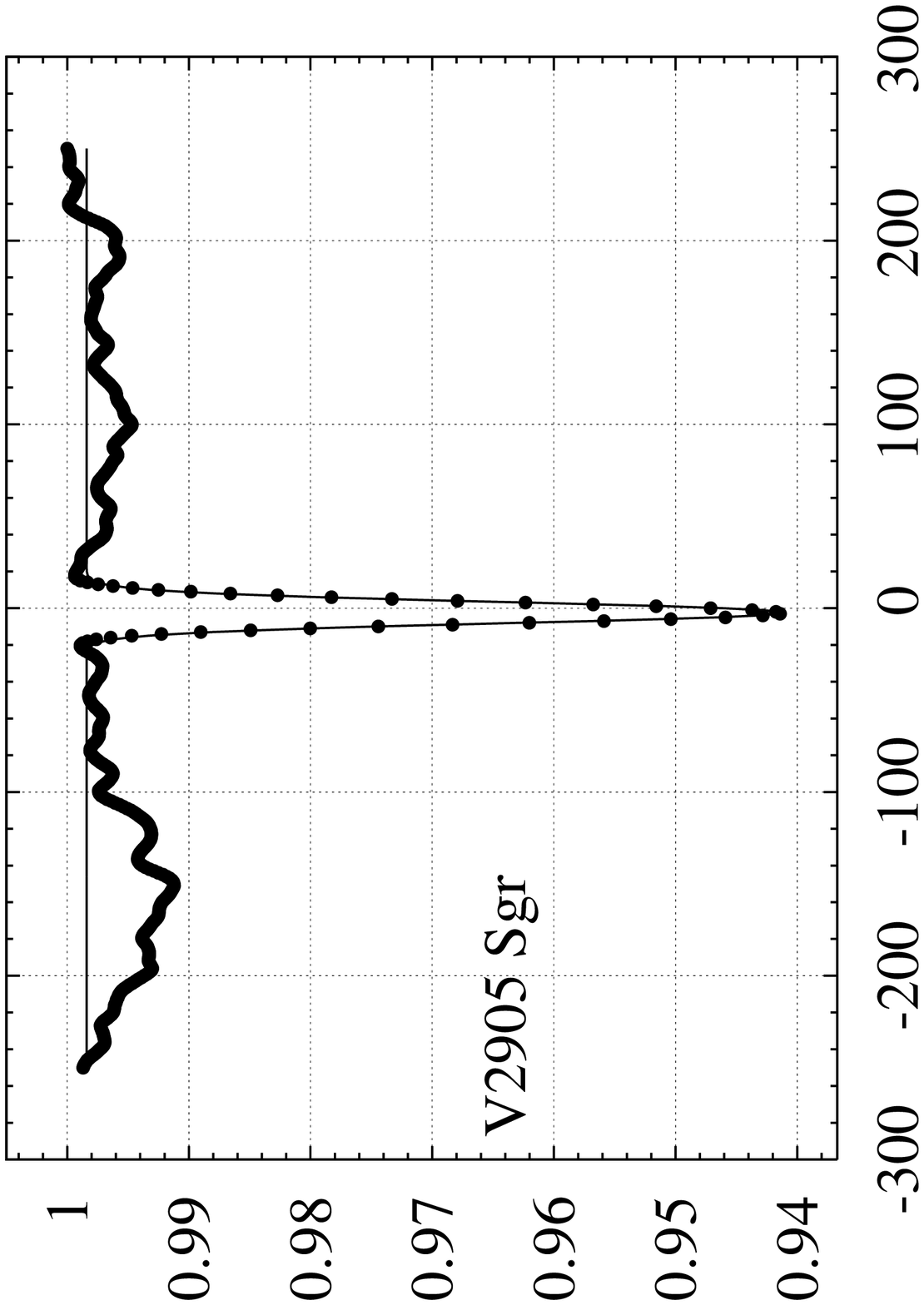}	
  \includegraphics{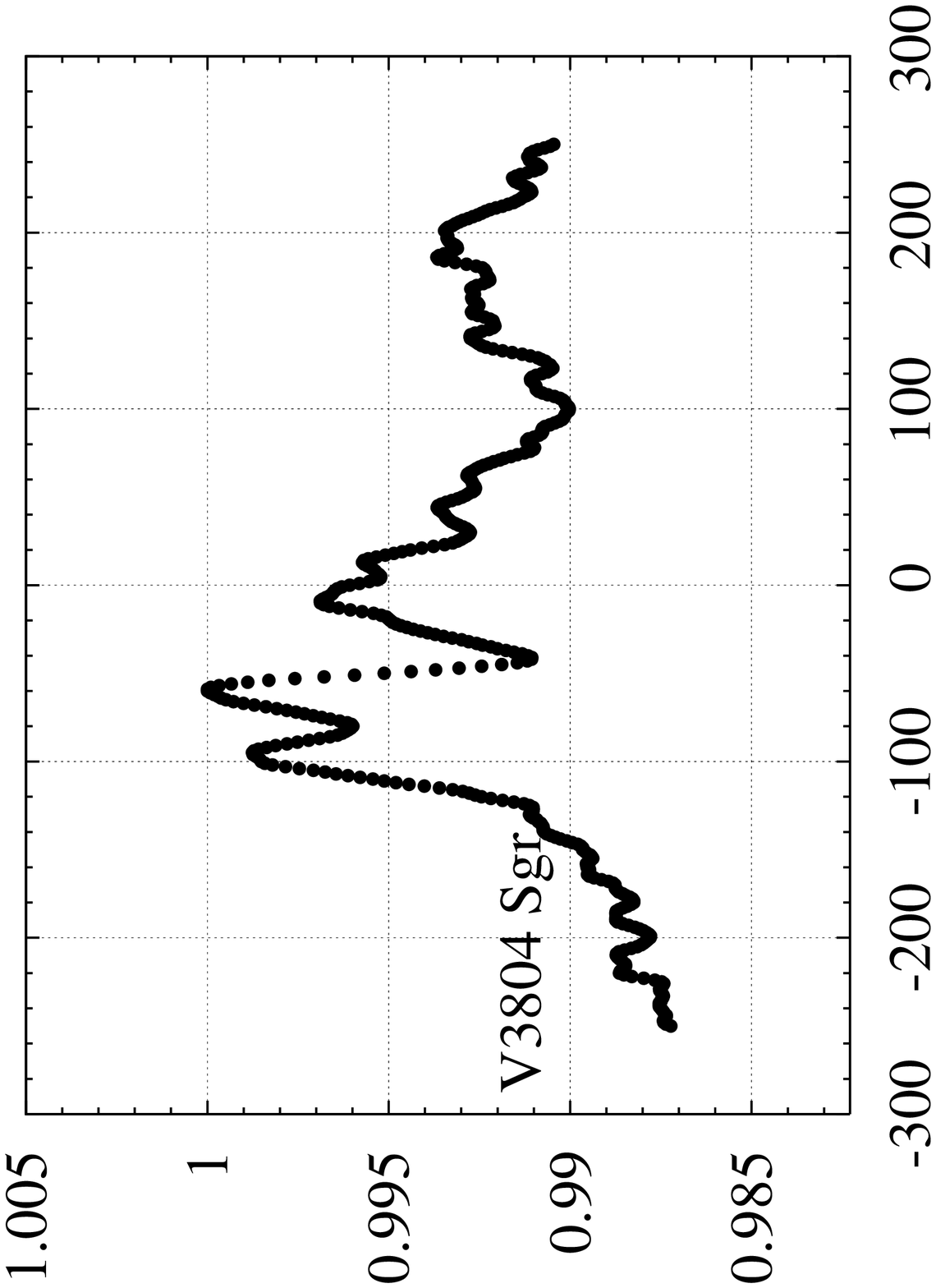}   
  \includegraphics{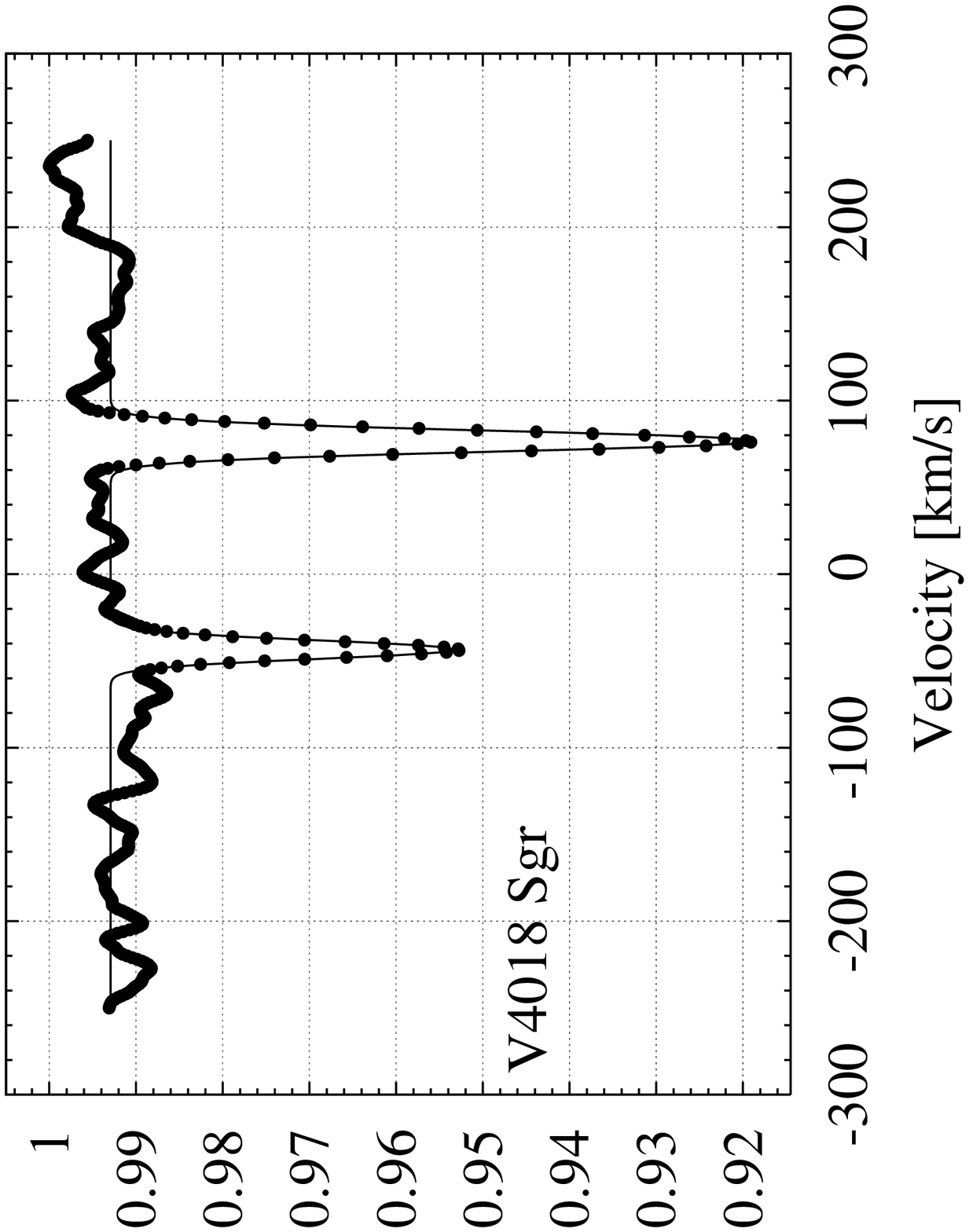}	
  \includegraphics{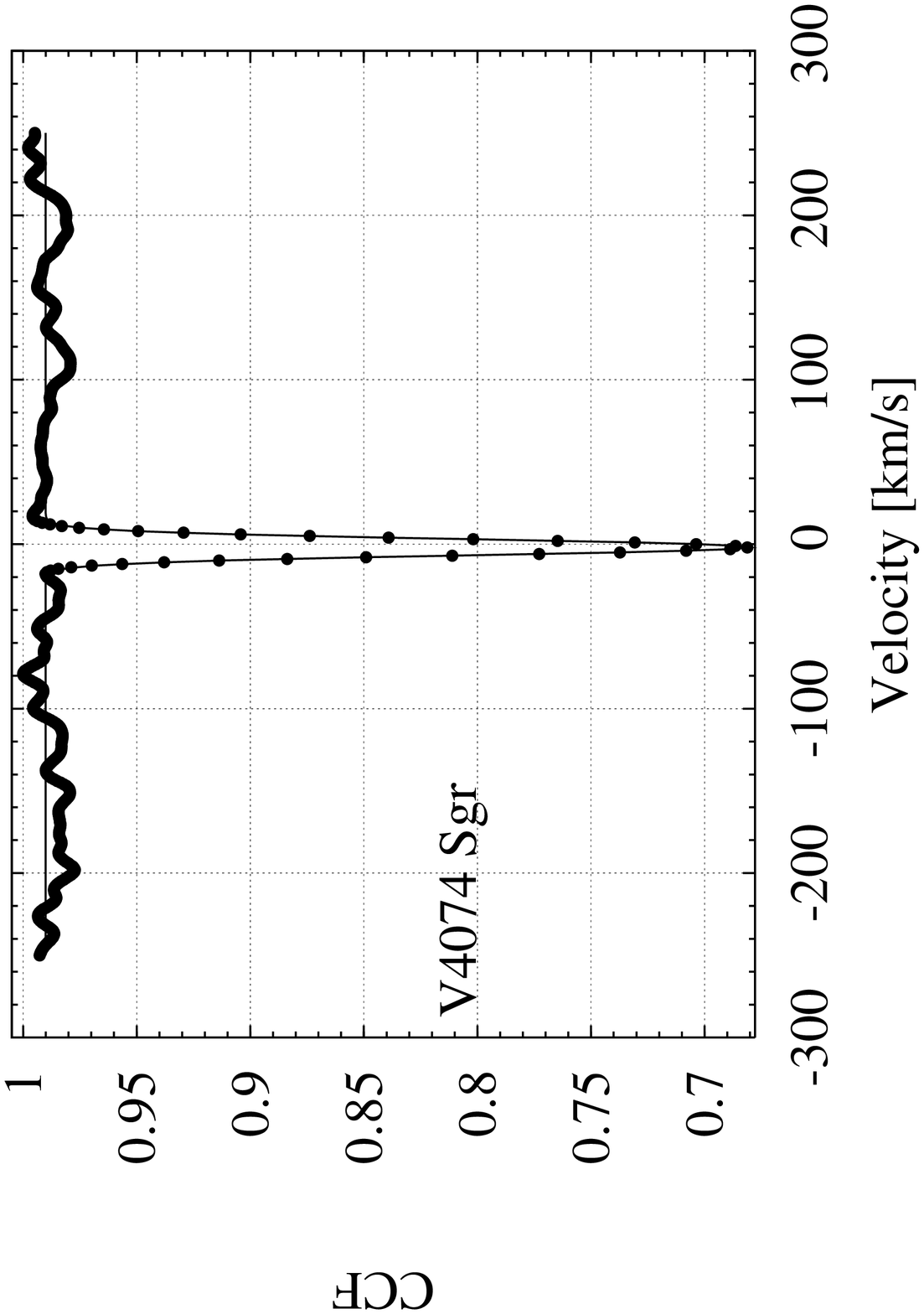}	
  \includegraphics{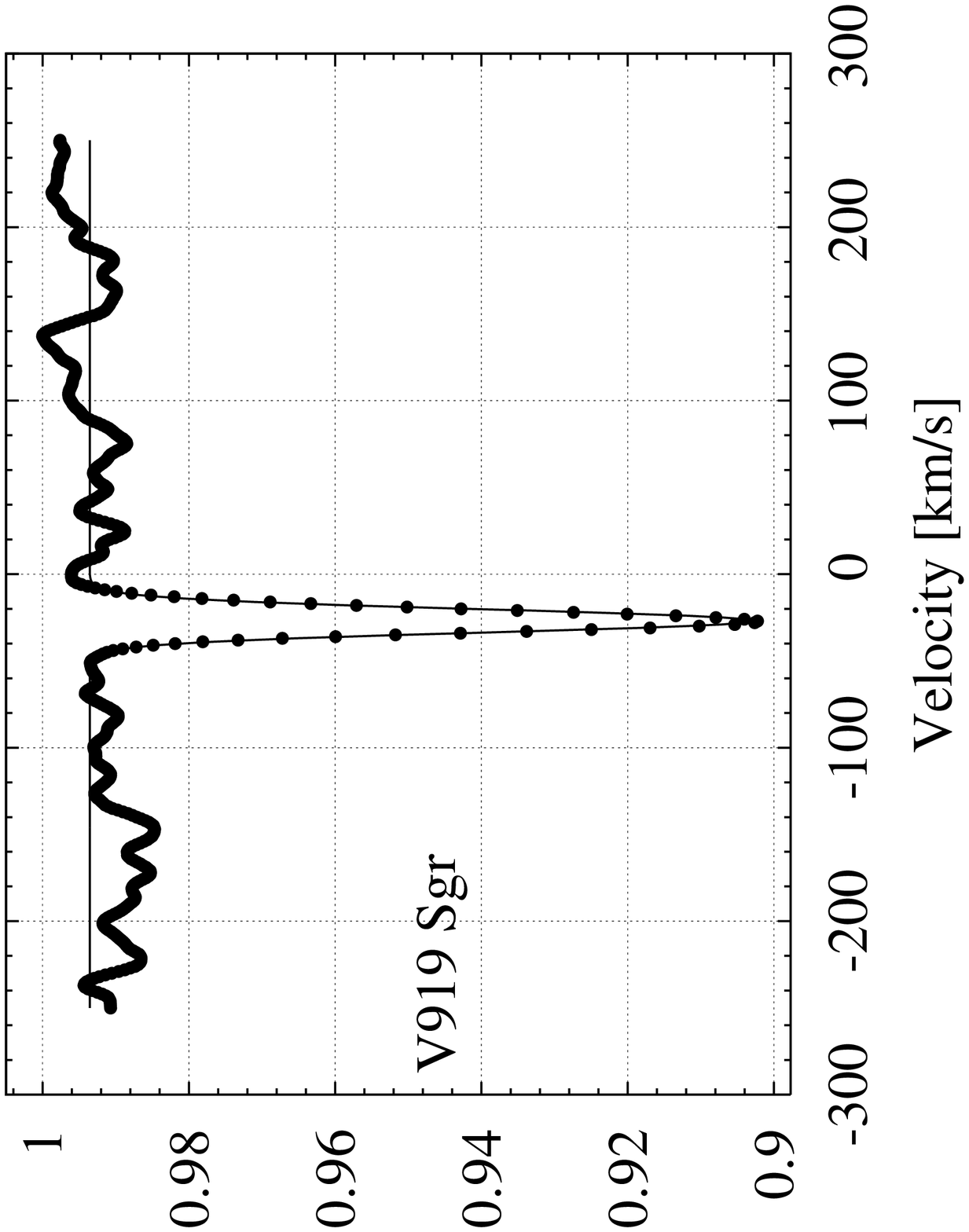}	
\caption[]{ continuation }	          
\end{figure*}


\begin{thebibliography}{99}  

\bibitem[Allen(1982)]{1982nss..coll...27A} Allen, D.~A.\ 1982, ASSL 
Vol.~95: IAU Colloq.~70: The Nature of Symbiotic Stars, 27 

\bibitem[Akritas \& Bershady(1996)]{1996ApJ...470..706A} Akritas, M.~G., \& 
Bershady, M.~A.\ 1996, ApJ, 470, 706 

\bibitem[Baranne et al.(1979)]{1979VA.....23..279B} Baranne, A., Mayor, M., 
\& Poncet, J.~L.\ 1979, Vistas in Astronomy, 23, 279 

\bibitem[Belczynski \& Mikolajewska(1998)]{1998MNRAS.296...77B} Belczynski, 
K., \& Mikolajewska, J.\ 1998, MNRAS, 296, 77 

\bibitem[Belczy{\' n}ski et al.(2000)]{2000A&AS..146..407B}   
Belczy{\'n}ski, K., Miko{\l}ajewska, J., Munari, U., Ivison, R.~J., 
\& Friedjung, M.,  2000,  A\&AS, 146, 407   


\bibitem[Bode et al.(2006)]{2006ApJ...652..629B} 
Bode, M. F., O'Brien, T. J., Osborne, J. P., et al. 2006,
ApJ, 652, 629 

\bibitem[Bruch et al.(1994)]{1994A&A...287..829B} Bruch, A., Niehues, M., 
\& Jones, A.~F.\ 1994, A\&A, 287, 829 

\bibitem[Chochol \& Wilson(2001)]{2001MNRAS.326..437C} Chochol, D., \& 
Wilson, R.~E.\ 2001, MNRAS, 326, 437 

\bibitem[Corradi et al.(2003)]{2003ASPC..303.....C} Corradi, R.~L.~M., 
Mikolajewska, J., \& Mahoney, T.~J.\ 2003, Symbiotic Stars Probing Stellar Evolution,
Astronomical Society of the  Pacific Conference Series, 303  

\bibitem[de Medeiros \& Mayor(1999)]{1999A&AS..139..433D} de Medeiros, 
J.~R., \& Mayor, M.\ 1999, A\&AS, 139, 433

\bibitem[Delfosse et al.(1998)]{1998A&A...331..581D} Delfosse, X., 
Forveille, T., Perrier, C., \& Mayor, M.\ 1998, A\&A, 331, 581 

\bibitem[Dobrzycka \& Kenyon(1994)]{1994AJ....108.2259D} Dobrzycka, D., 
\& Kenyon, S.~J.\ 1994, AJ, 108, 2259 

\bibitem[Dobrzycka et al.(1996)]{1996AJ....111.2090D} Dobrzycka, D., 
Kenyon, S.~J., Proga, D., Mikolajewska, J., \& Wade, R.~A.\ 1996, AJ, 111, 2090 

\bibitem[Dumm et al.(1998)]{1998A&A...336..637D} Dumm, T., Muerset, U., 
Nussbaumer, H., Schild, H., Schmid, H.~M., Schmutz, W., \& Shore, S.~N.\ 
1998, A\&A, 336, 637 

\bibitem[Dumm \& Schild(1998)]{1998NewA....3..137D} Dumm, T., \& Schild, 
H.\ 1998, New Astronomy, 3, 137 


\bibitem[Fekel(1981)]{1981ApJ...246..879F} Fekel, F.~C., Jr.\ 1981, ApJ, 246, 879 

\bibitem[Fekel(1997)]{1997PASP..109..514F} Fekel, F.~C.\ 1997, PASP, 109, 514

\bibitem[Fekel et al.(2000)]{2000AJ....119.1375F} Fekel, F.~C., Joyce, 
R.~R., Hinkle, K.~H., \& Skrutskie, M.~F.\ 2000a, AJ, 119, 1375 

\bibitem[Fekel et al.(2000)]{2000AJ....120.3255F} Fekel, F.~C., Hinkle, 
K.~H., Joyce, R.~R., \& Skrutskie, M.~F.\ 2000b, AJ, 120, 3255 

\bibitem[Fekel et al.(2001)]{2001AJ....121.2219F} Fekel, F.~C., Hinkle, 
K.~H., Joyce, R.~R., \& Skrutskie, M.~F.\ 2001, AJ, 121, 2219 

\bibitem[Friedjung et al.(2003)]{2003A&A...400..595F} Friedjung, M., 
G{\'a}lis, R., Hric, L., \& Petr{\'{\i}}k, K.\ 2003, A\&A, 400, 595 

\bibitem[Fekel et al.(2004)]{2004IAUS..215..168F} Fekel, F.~C., Hinkle, 
K.~H., \& Joyce, R.~R.\ 2004, IAU Symposium, 215, 168 

\bibitem[Fekel et al.(2007)]{2007AJ....133...17F} Fekel, F.~C., Hinkle, 
K.~H., Joyce, R.~R., Wood, P.~R., \& Lebzelter, T.\ 2007, AJ, 133, 17

\bibitem[Gray(1976)]{1976oasp.book.....G} Gray, D.~F.\ 1976, 
The observation and analysis of stellar photospheres, Research 
supported by the National Research Council of Canada.~New York, 
Wiley-Interscience, 1976.~484 p.

\bibitem[Hale(1994)]{1994AJ....107..306H} Hale, A.\ 1994, AJ, 107, 306 

\bibitem[Harries \& Howarth(1996)]{1996A&A...310..235H} Harries, T.~J., \& 
Howarth, I.~D.\ 1996, A\&A, 310, 235 

\bibitem[Harries \& Howarth(2000)]{2000A&A...361..139H} Harries, T.~J., \& 
Howarth, I.~D.\ 2000, A\&A, 361, 139 

\bibitem[Henry et al.(2000)]{2000ApJS..130..201H} Henry, G.~W., Fekel,
F.~C., Henry, S.~M., \& Hall, D.~S.\ 2000, ApJS, 130, 201

\bibitem[Hoffleit(1970)]{1970IBVS..469....1H} Hoffleit, D.\ 1970, 
Informational Bulletin on Variable Stars, 469, 1 

\bibitem[]{} 
Hubeny I., Lanz T., Jeffery C.S., 1994, in Jeffery C.S., eds, 
Newsletter on Analysis of Astronomical Spectra, No.20, CCP7 St. Andrews Univ., St. Andrews, p.30

\bibitem[Hut(1981)]{1981A&A....99..126H} Hut, P.\ 1981, A\&A, 99, 126 


\bibitem[Kaufer et al.(1999)]{1999Msngr..95....8K} Kaufer, A., Stahl, O.,  
Tubbesing, S., Norregaard, P., Avila, G., Francois, P., Pasquini, L., \&  
Pizzella, A.\ 1999, The Messenger, 95, 8  
 
\bibitem[Kenyon(1986)]{1986syst.book.....K} Kenyon, S.~J.\ 1986, 
The symbiotic stars, Cambridge and New York, Cambridge University Press, 1986, 295 p.,  

\bibitem[Kenyon \& Fernandez-Castro(1987)]{1987AJ.....93..938K} Kenyon, 
S.~J., \& Fernandez-Castro, T.\ 1987, AJ, 93, 938 

\bibitem[Kenyon \& Garcia(1989)]{1989AJ.....97..194K} Kenyon, S.~J., \& 
Garcia, M.~R.\ 1989, AJ, 97, 194 

\bibitem[Kenyon \& Mikolajewska(1995)]{1995AJ....110..391K} Kenyon, S.~J., 
\& Mikolajewska, J.\ 1995, AJ, 110, 391 

\bibitem[Keyes \& Preblich(2004)]{2004AJ....128.2981K} Keyes, C.~D., 
\& Preblich, B.\ 2004, AJ, 128, 2981 

\bibitem[Kenyon et al.(1993)]{1993AJ....106.1573K} Kenyon, S.~J., 
Mikolajewska, J., Mikolajewski, M., Polidan, R.~S., \& Slovak, M.~H.\ 1993, 
AJ, 106, 1573 

\bibitem[]{} Kupka F., Piskunov N.E., 
Ryabchikova TA., Stempels H.C., Weiss W.W., 1999, A\&AS, 138, 119


\bibitem[Kurucz(1993)]{1993...Kurucz....ATLAS9} 
Kurucz R.L., 1993, ATLAS9 Stellar Atmosphere Programs (Kurucz CD-ROM 13)

\bibitem[Medina Tanco \& Steiner(1995)]{1995AJ....109.1770M} Medina Tanco, 
G.~A., \& Steiner, J.~E.\ 1995, AJ, 109, 1770 

\bibitem[Melo et al.(2001)]{2001A&A...375..851M} Melo, C.~H.~F., Pasquini, 
L., \& De Medeiros, J.~R.\ 2001, A\&A, 375, 851 

\bibitem[Melo(2003)]{2003A&A...410..269M} Melo, C.~H.~F.\ 2003, A\&A, 410, 269  

\bibitem[Mikolajewska et al.(1989)]{1989AJ.....98.1427M} Mikolajewska, J., 
Mikolajewski, M., \& Kenyon, S.~J.\ 1989, AJ, 98, 1427  

\bibitem[Mikolajewska et al.(1995)]{1995AJ....109.1289M} Mikolajewska, J., 
Kenyon, S.~J., Mikolajewski, M., Garcia, M.~R., \& Polidan, R.~S.\ 1995, 
AJ, 109, 1289 

\bibitem[Miko{\l}ajewska(2003)]{2003ASPC..303....9M} Miko{\l}ajewska, J.\ 
2003, Astronomical Society of the Pacific Conference Series, 303, 9 

\bibitem[M{\" u}rset \& Schmid(1999)]{1999A&AS..137..473M} M{\" u}rset, U., 
\& Schmid, H.~M.\ 1999, A\&AS, 137, 473 

\bibitem[M{\" u}rset et al.(2000)]{2000A&A...353..952M} M{\" u}rset, U., 
Dumm, T., Isenegger, S., Nussbaumer, H., Schild, H., Schmid, H.~M., \& 
Schmutz, W.\ 2000, A\&A, 353, 952 

\bibitem[Nordstr{\"o}m et al.(2004)]{2004A&A...418..989N} Nordstr{\"o}m,
B., et al.\ 2004, A\&A, 418, 989

\bibitem[Pereira et al.(1995)]{1995A&A...293..783P} Pereira, C.~B., Vogel, 
M., \& Nussbaumer, H.\ 1995, A\&A, 293, 783 

\bibitem[Pereira et al.(2005)]{2005A&A...429..993P} Pereira, C.~B., Smith, 
V.~V., \& Cunha, K.\ 2005, A\&A, 429, 993 


\bibitem[Pucinskas(1970)]{1970VilOB..27...24P} Pucinskas, A.\ 1970, Vilnius 
Astronomijos Observatorijos Biuletenis, 27, 24 

\bibitem[Quiroga et al.(2002)]{2002A&A...387..139Q} 
Quiroga, C., Miko{\l}ajewska, J., Brandi, E., Ferrer, O., \& Garc{\'{\i}}a, L.\ 
2002, A\&A, 387, 139 

\bibitem[Schild et al.(1996)]{1996A&A...306..477S} Schild, H., Muerset, U., 
\& Schmutz, W.\ 1996, A\&A, 306, 477 

\bibitem[Schild et al.(2001)]{2001A&A...366..972S} Schild, H., Dumm, T., 
M{\" u}rset, U., Nussbaumer, H., Schmid, H.~M., \& Schmutz, W.\ 2001, A\&A, 
366, 972 


\bibitem[Schmid et al.(1998)]{1998A&A...329..986S} Schmid, H.~M., Dumm, T., 
Murset, U., Nussbaumer, H., Schild, H., \& Schmutz, W.\ 1998, A\&A, 329, 
986 

\bibitem[Schmidt-Kaler (1982)]{1982book....}  
Schmidt-Kaler, T. H. 1982, in Landolt-Börnstein, 
New Series, Group VI, Vol. 2b, Stars and Star Clusters, 
ed. K. Schaifers \& H. H. Voigt (New York: Springer) 


\bibitem[Schmutz et al.(1994)]{1994A&A...288..819S} Schmutz, W., Schild, 
H., Muerset, U., \& Schmid, H.~M.\ 1994, A\&A, 288, 819 

\bibitem[Skopal et al.(1997)]{1997MNRAS.292..703S} Skopal, A., Vittone, A., 
Errico, L., Bode, M.~F., Lloyd, H.~M., \& Tamura, S.\ 1997, MNRAS, 292, 703 

\bibitem[Skopal(2005)]{2005A&A...440..995S} Skopal, A.\ 2005, A\&A, 440, 
995 

\bibitem[Smith et al.(1997)]{1997A&A...324...97S} Smith, V.~V., Cunha, K., 
Jorissen, A., \& Boffin, H.~M.~J.\ 1997, A\&A, 324, 97 

\bibitem[Smith et al.(2001)]{2001ApJ...556L..55S}
Smith, V.~V., Pereira, C.~B., \& Cunha, K.\ 2001, ApJ, 556, L55

\bibitem[Soker(2002)]{2002MNRAS.337.1038S} Soker, N.\ 2002, MNRAS, 337, 
1038 

\bibitem[Stanishev et al.(2004)]{2004A&A...415..609S} Stanishev, V., 
Zamanov, R., Tomov, N., \& Marziani, P.\ 2004, A\&A, 415, 609 

\bibitem[Stawikowski(1994)]{1994AcA....44..393S} Stawikowski, A.\ 1994, 
Acta Astronomica, 44, 393 

\bibitem[Tassoul(2000)]{2000stro.book.....T} Tassoul, J.\ 2000, Stellar  
rotation / Jean-Louis Tassoul.~Cambridge ; New York : Cambridge University  
Press, 2000 ~(Cambridge astrophysics series  36)   

\bibitem[Tomov et al.(2000)]{2000A&A...364..557T} Tomov, N.~A., Tomova, 
M.~T., \& Ivanova, A.\ 2000, A\&A, 364, 557 

\bibitem[van Belle et al.(1999)]{1999AJ....117..521V} van Belle, G.~T., Lane, B.F., 
Thompson, R.R., Doden, A.F., Colavita, M.M.,  et  al.\ 1999, AJ, 117, 521 



\bibitem[Yudin \& Munari(1993)]{1993A&A...270..165Y} Yudin, B., \& Munari, 
U.\ 1993, A\&A, 270, 165 

\bibitem[Yudin et al.(2005)]{2005ARep...49..232Y} Yudin, B.~F., Shenavrin, 
V.~I., Kolotilov, E.~A., Tatarnikova, A.~A., \& Tatarnikov, A.~M.\ 2005, 
Astronomy Reports, 49, 232 

\bibitem[Yungelson et al.(1995)]{1995ApJ...447..656Y} Yungelson, L., Livio, 
M., Tutukov, A., \& Kenyon, S.~J.\ 1995, ApJ, 447, 656 

\bibitem[Zahn(1977)]{1977A&A....57..383Z} Zahn, J.-P.\ 1977, A\&A, 57, 383  
 
\bibitem[Zahn(1989)]{1989A&A...220..112Z} Zahn, J.-P.\ 1989, A\&A, 220, 112  


\bibitem[Zamanov et al.(2006)]{2006MNRAS.365.1215Z} Zamanov, R.~K., Bode, 
M.~F., Melo, C.~H.~F., Porter, J., Gomboc, A., \& Konstantinova-Antova, R.\ 
2006, MNRAS, 365, 1215 (Paper~I) 

\bibitem[Zhu et al.(1999)]{1999A&AS..140...69Z} Zhu, Z.~X., Friedjung, M., 
Zhao, G., Hang, H.~R., \& Huang, C.~C.\ 1999, A\&AS, 140, 69 

\end{thebibliography}
\end{document}